\documentclass[%
 aip,
 jcp,
 amsmath,amssymb,
reprint,%
]{revtex4-1}

\usepackage[utf8]{inputenc}
\usepackage[T1]{fontenc}
\usepackage{mathptmx}
\usepackage{etoolbox}
\usepackage{graphicx}  % Include figure files
\usepackage{bm}        % Bold
\usepackage{bbm}       % Bold
\usepackage{physics}   % Bra, Ket, and related stuff
\usepackage{textgreek} % Write greek letters inb text mode
\usepackage{subfigure} % Subfigures
\usepackage{multirow}  % Multicolumns and rows for tables
\usepackage{booktabs}  % Nice tables
\usepackage{hyperref}  % References in PDF
\usepackage{longtable} % for long tables spanning multiple pages
\usepackage{siunitx}   % SI-units, table formatting
\usepackage[12pt]{moresize}

\usepackage{pdfpages} % include pdfs
\usepackage{pgffor} % for loops
\usepackage{xr} % referencing the supplement

\def\supplementfilename{Supplementary-Material}

\pdfximage{\supplementfilename.pdf}
\def\numbersupplementpages{\the\pdflastximagepages}

\newif\ifarXiv
\arXivtrue 
\makeatletter
\AtBeginDocument{\let\LS@rot\@undefined}
\makeatother

\makeatletter
\def\@email#1#2{%
 \endgroup
 \patchcmd{\titleblock@produce}
  {\frontmatter@RRAPformat}
  {\frontmatter@RRAPformat{\produce@RRAP{*#1\href{mailto:#2}{#2}}}\frontmatter@RRAPformat}
  {}{}
}%
\makeatother
\begin{document}

%%%%%%%%%%%%%%%%%%%%%%%%%%%%%%%%%%%%%%%%%%%%%%%%%%%%%%%%%%%%%%%%%%%%%
%% Authors and title
%%%%%%%%%%%%%%%%%%%%%%%%%%%%%%%%%%%%%%%%%%%%%%%%%%%%%%%%%%%%%%%%%%%%%
\title[CDFT for Spin--Orbit Coupling: Extension to Periodic Systems]
{Current density functional framework for spin--orbit coupling: Extension to periodic systems}

\author{Yannick J. Franzke}
\affiliation{Otto Schott Institute of Materials Research,
Friedrich Schiller University Jena, L{\"o}bdergraben 32, 07743 Jena, Germany}
\email[Email for correspondence: ]{yannick.franzke@uni-jena.de}
%ORCID: 0000-0002-8344-113X

\author{Christof Holzer}
\affiliation{Institute of Theoretical Solid State Physics, Karlsruhe
Institute of Technology (KIT), Wolfgang-Gaede-Stra\ss{}e 1, 76131 Karlsruhe,
Germany}
\email[Email for correspondence: ]{christof.holzer@kit.edu}
%ORCID: 0000-0001-8234-260X

\date{\today}% It is always \today, today,

%%%%%%%%%%%%%%%%%%%%%%%%%%%%%%%%%%%%%%%%%%%%%%%%%%%%%%%%%%%%%%%%%%%%%
%% Abstract and cover page
%%%%%%%%%%%%%%%%%%%%%%%%%%%%%%%%%%%%%%%%%%%%%%%%%%%%%%%%%%%%%%%%%%%%%
\begin{abstract}
Spin--orbit coupling induces a current density in the ground
state, which consequently requires a generalization for
\textit{meta}-generalized gradient approximations.
That is, the exchange-correlation energy has to be constructed as
an explicit functional of the current density and a generalized
kinetic energy density has to be formed to satisfy theoretical
constraints. Herein, we generalize our previously presented formalism
of spin--orbit current density functional theory [Holzer \textit{et al.},
J.\ Chem.\ Phys.\ \textbf{157}, 204102 (2022)] to non-magnetic
and magnetic periodic systems of arbitrary dimension. Besides
the ground-state exchange-correlation potential, analytical
derivatives such as geometry gradients and stress tensors are
implemented.
The importance of the current density is assessed for band gaps,
lattice constants, magnetic transitions, and Rashba splittings.
For the latter, the impact of the current density may be larger than
the deviation between different density functional approximations.
\end{abstract}

\maketitle

%% Communications have a length limit of 3500 words.
%% Note that figures, equations, tables and references should
%% not be included in the 3500 word count.
%%%%%%%%%%%%%%%%%%%%%%%%%%%%%%%%%%%%%%%%%%%%%%%%%%%%%%%%%%%%%%%%%%%%%
\section{Introduction}
\label{sec:introduction}
In the constantly evolving field of density functional theory (DFT), 
especially the construction of \textit{meta}-generalized 
gradient approximations (meta-GGAs) has received great attention 
over the last two decades. \cite{Burke:Perspective.2012, Becke:Perspective.2014, 
Sun.Ruzsinszky.ea:Strongly.2015, Mardirossian.Head-Gordon:Thirty.2017}
Modern meta-GGAs use the iso-orbital constraint and the 
von-Weiz\"acker inequality to identify one-electron regions and, thus 
cancelling self-interaction errors in single electron regions.
\cite{Sun.Ruzsinszky.ea:Strongly.2015,
Tao.Mo:Accurate.2016} According to benchmark calculations,
\cite{Mardirossian.Head-Gordon:Thirty.2017, Hao.Sun.ea:Performance.2013, 
Mo.Car.ea:Assessment.2017, Goerigk.Hansen.ea:look.2017,
Franzke.Yu:Hyperfine.2022, Franzke.Yu:Quasi-Relativistic.2022,
Becke:Density-Functional.2022, Borlido.Aull.ea:Large-Scale.2019,
Holzer.Franzke.ea:Assessing.2021, Holzer.Franzke:Local.2022,
Lee.Feng.ea:Approaching.2021, Ghosh.Jana.ea:Efficient.2022,
Kovacs.Tran.ea:What.2022, Liang.Feng.ea:Revisiting.2022,
Franzke:Reducing.2023, Kovacs.Blaha.ea:Origin.2023,
Lebeda.Aschebrock.ea:Right.2023}
meta-GGAs outperform the preceding generalized gradient
approximations (GGAs) at roughly the same computational cost.
However, external magnetic fields \cite{Furness.Verbeke.ea:Current.2015,
Tellgren.Teale.ea:Non-perturbative.2014, Irons.David.ea:Optimizing.2021, 
Pausch.Holzer:Linear.2022} or spin--orbit coupling
\cite{InB-Saue:2005, Saue:Relativistic.2011, Pyykko:Relativistic.2012}
necessitate further generalizations for meta-GGAs to meet
theoretical constrains, as the current density alters the
curvature of the Fermi hole in its second-order Taylor expansion.
\cite{Dobson:Alternative.1993, Tao:Explicit.2005,
Pittalis.Kurth.ea:Orbital.2007, Rasanen.Pittalis.ea:Universal.2010,
Pittalis.Rasanen.ea:Gaussian.2009, Bates.Furche:Harnessing.2012,
Maier.Ikabata.ea:Relativistic.2020, Holzer.Franke.ea:Current.2022,
Desmarais.Ambrogio.ea:Generalized.2024, Desmarais.Maul.ea:2024}
Density functional approximations (DFAs) constructed by
taking into account this change in curvature are termed
``CDFT'' functionals. \cite{Tellgren.Kvaal.ea:Choice.2012,
Furness.Verbeke.ea:Current.2015} CDFT based approximations
are for example necessary for the iso-orbital indicator
to remain bounded between 0 and 1.
\cite{Holzer.Franke.ea:Current.2022, Maier.Ikabata.ea:Relativistic.2020}
In external magnetic fields, this correction is even necessary
to guarantee gauge-invariance.
\cite{Furness.Verbeke.ea:Current.2015, Pausch.Holzer:Linear.2022,
Holzer.Franke.ea:Current.2022}

Furthermore, certain current-carrying ground states also
heavily depend on the correct introduction of the current 
density, with a failure to account for them leading
to large deviation for meta-GGAs.
\cite{Becke:Current.2002, Holzer.Franke.ea:Current.2022} 
We recently presented a current-dependent formulation of
density functional theory for spin--orbit coupling in the 
molecular regime. \cite{Holzer.Franke.ea:Current.2022}
Here, inclusion of the current density
leads to notable changes for both closed-shell Kramers-restricted
(KR) and open-shell Kramers-unrestricted (KU) calculations.

To account for the change in Fermi hole curvature, 
the kinetic energy density $\tau$ is generalized with the current
density $\Vec{j}$. In a two-component (2c) non-collinear
formalism, \cite{Kubler.Hock.ea:Density.1988, Van-Wullen:Spin.2002,
Saue.Helgaker:Four-component.2002, Armbruster.Weigend.ea:Self-consistent.2008,
Peralta.Scuseria.ea:Noncollinear.2007, Scalmani.Frisch:New.2012,
Bulik.Scalmani.ea:Noncollinear.2013, Baldes.Weigend:Efficient.2013,
Egidi.Sun.ea:Two-Component.2017, Komorovsky.Cherry.ea:Four-component.2019, 
Desmarais.Komorovsky.ea:Spin-orbit.2021}
this results in the generalized current density
\begin{equation}
    \tilde{\tau}_{\uparrow,\downarrow} = \tau_{\uparrow,\downarrow}  - \frac{|\vec{j}_{\uparrow,\downarrow}|^{\,2}}{2 n_{\uparrow,\downarrow}}
\end{equation}
based on the spin-up and down quantities. \cite{Holzer.Franke.ea:Current.2022}
These are formed with the particle and spin-magnetization contributions,
e.g. the spin-up and down electron density $n_{\uparrow,\downarrow}$
follows as
\begin{equation}
n_{\uparrow,\downarrow} (\vec{r}) = \frac{1}{2} \left[ n(\vec{r}) \pm |\vec{m}(\vec{r})| \right]  = \frac{1}{2} \left[ n(\vec{r}) \pm s (\vec{r}) \right]
\end{equation}
with the particle density $n$, the spin-magnetization vector $\vec{m}$,
and the spin density $s$.
Therefore, the exchange-correlation (XC) energy of a ``pure'' functional
\cite{Becke:Perspective.2014} explicitly depends on the current density
\begin{equation}
\begin{split}
E^{\text{XC}} & = \int f^{\text{XC}} \left[
n_{\uparrow,\downarrow}(\vec{r}), 
\gamma_{\uparrow \uparrow,\uparrow \downarrow,\downarrow \downarrow}(\vec{r}),
\tau_{\uparrow,\downarrow}(\vec{r}), \vec{j}_{\uparrow, \downarrow}(\vec{r})\right] ~ \textrm{d}^3 r \\
 & = \int g^{\text{XC}} \left[n_{\uparrow,\downarrow}(\vec{r}), 
\gamma_{\uparrow \uparrow,\uparrow \downarrow,\downarrow \downarrow}(\vec{r}),
\tilde{\tau}_{\uparrow,\downarrow}(\vec{r})\right] ~ \textrm{d}^3 r
\end{split}
\end{equation}
where $f^{\text{XC}}$ describes the density functional approximation
and $\gamma_{\uparrow \downarrow} (\vec{r}) = \frac{1}{4} \left( \vec{\nabla} n_{\uparrow}(\vec{r}) \right) \cdot \left( \vec{\nabla} n_{\downarrow}(\vec{r}) \right)$,
hence $\gamma_{\uparrow \downarrow} = \gamma_{\downarrow \uparrow}$.
For a Kramers-restricted system, the spin-magnetization vector and
the particle current density vanish. However, the spin-current
density is generally non-zero and thus it is still necessary to
form the generalized kinetic energy density.

In this work, we will extend our previous CDFT formulation
\cite{Holzer.Franke.ea:Current.2022} to non-magnetic and magnetic
periodic systems. We assess the importance of the current density
for band gaps, cell structures, magnetic transitions, and Rashba
splittings with common meta-GGAs. Together with previous studies
on the impact of spin--orbit-induced current densities for
meta-GGAs \cite{Holzer.Franke.ea:Current.2022, Bruder.Franzke.ea:Paramagnetic.2022,
Bruder.Franzke.ea:Zero-Field.2023, Franzke.Bruder.ea:Paramagnetic.2024,
Desmarais.Maul.ea:2024}
this will further help to set guidelines and recommendations for
the application of CDFT to the different properties and functionals
for discrete and periodic systems.

\section{Theory}
\label{sec:theory}
The one-particle density matrix associated with the two-component
spinor functions $\vec{\psi}_i{^{\vec{k}}}$ at a given $\vec{k}$ point
reads
\begin{equation}
\begin{split}
    n(\vec{r},\vec{r}\,') =  & \frac{1}{V_{\textrm{FBZ}}} \sum_{i=1}^n \int_{\textrm{FBZ}}^{\epsilon_i^{\vec{k}} < \epsilon_{\textrm{F}}} \vec{\psi}_i{^{\vec{k}}}(\vec{r})\left(\vec{\psi}_i{^{\vec{k}}}(\vec{r}\,')\right)^\dagger \textrm{d}^3k \\
    = & \left(\begin{array}{cc} n^{\alpha \alpha}(\vec{r},\vec{r}\,') & n^{\alpha \beta}(\vec{r},\vec{r}\,') \\
n^{\beta \alpha}(\vec{r},\vec{r}\,') & n^{\beta \beta}(\vec{r},\vec{r}\,') \end{array}\right)
\end{split}
\end{equation}
where $V_{\textrm{FBZ}}$ is the volume of the first Brillouin zone (FBZ),
$\epsilon_i$ the energy eigenvalue, and $\epsilon_{\textrm{F}}$ the
Fermi level. The spinor functions are expanded with Bloch functions,
$\phi_\mu$, based on atomic orbital (AO) basis functions, $\chi_{\mu}$, as
\begin{eqnarray}
\allowdisplaybreaks
\vec{\psi}_i{^{\vec{k}}}(\vec{r}) & = & \begin{pmatrix} \psi_i^{\alpha, \vec{k}}(\vec{r}) \\ \psi_i^{\beta, \vec{k}}(\vec{r})\end{pmatrix}
= \sum_{\mu} \begin{pmatrix} c_{\mu i}^{\alpha, \vec{k}} \\ c_{\mu i}^{ \beta, \vec{k}} \end{pmatrix}
\phi_\mu^{\vec{k}}(\vec{r}) \\
\phi_\mu^{\vec{k}}(\vec{r}) & = & \frac{1}{\sqrt{N_{\text{UC}}}}
\sum_{\vec{L}}e^{\text{i}\vec{k}\cdot\vec{L}} ~ \chi_\mu^{\vec{L}}(\vec{r}) \textrm{.}
\label{eq:bloch}
\end{eqnarray}
$N_{\text{UC}}$ denotes the number of electrons in the unit cell (UC)
and $\vec{L}$ the lattice vector.
Thus, all density variables are available from the AO density matrix in
position space given by
\begin{align}
D_{\mu \nu}^{\sigma \sigma', \vec{L} \vec{L}\,'}
 = \frac{1}{V_{\textrm{FBZ}}} \sum_{i} \int_{\textrm{FBZ}}^{\epsilon_i^{\vec{k}} < \epsilon_{\textrm{F}}} 
e^{\text{i} \vec{k}\cdot \left[\vec{L} - \vec{L}\,'\right]}
\left(c_{\mu i}^{\sigma,\vec{k}} ~ c_{\nu i}^{*\sigma',\vec{k}} \right) \text{d}^3k
\end{align}
with the expansion coefficients $c_{\mu i}$ and $\sigma, \sigma' \in \{\alpha, \beta\}$.
The complete two-component AO density matrix reads
\begin{equation}
\textbf{D}^{\vec{L} \vec{L}\,'} = \left(\begin{array}{cc} \textbf{D}^{\alpha \alpha} & \textbf{D}^{\alpha \beta} \\
\textbf{D}^{\beta \alpha} & \textbf{D}^{\beta \beta} \end{array}\right)^{\vec{L} \vec{L}\,'}
\quad \text{with} \quad \left(\textbf{D}^{\vec{L} \vec{L}\,'} \right)^{\dagger} = 
\textbf{D}^{\vec{L}\,' \vec{L}} \textrm{.}
\end{equation}
In the spirit of Bulik \textit{et al.}, \cite{Bulik.Scalmani.ea:Noncollinear.2013}
the real symmetric (RS), real antisymmetric (RA), imaginary symmetric (IS), and
imaginary antisymmetric (OA) linear combinations
\begin{eqnarray}
\left[ \textbf{D}_{\text{RS, RA}}^{\sigma \sigma'} \right]^{\vec{L} \vec{L}\,'} & = &
\frac{1}{2} \left[ \text{Re} \left( \textbf{D}^{\sigma \sigma'} \pm \textbf{D}^{\sigma' \sigma}  \right)
\right]^{\vec{L} \vec{L}\,'} \\
\left[ \textbf{D}_{\text{IA, IS}}^{\sigma \sigma'} \right]^{\vec{L} \vec{L}\,'} & = &
\frac{1}{2} \left[ \text{Im} \left(  \textbf{D}^{\sigma \sigma'} \pm \textbf{D}^{\sigma' \sigma}   \right)
\right]^{\vec{L} \vec{L}\,'}
\end{eqnarray}
are formed. Of course, the same-spin antisymmetric contributions are
zero. The electron density and its derivatives are available
from the symmetric linear combinations
\begin{eqnarray}
\allowdisplaybreaks
n(\vec{r}) & = & \sum_{\mu \nu} \sum_{\vec{L} \vec{L}\,'} \left[\textbf{D}_{\text{RS}}^{\alpha \alpha} + \textbf{D}_{\text{RS}}^{\beta \beta}  \right]_{\mu \nu}^{\vec{L} \vec{L}\,'} \chi_{\mu}^{\vec{L}} (\vec{r}) ~ \chi_{\nu}^{\vec{L}\,'} (\vec{r}) \\
m_x(\vec{r}) & = & \sum_{\mu \nu} \sum_{\vec{L} \vec{L}\,'} 2 \left[\textbf{D}_{\text{RS}}^{\alpha \beta} \right]_{\mu \nu}^{\vec{L} \vec{L}\,'} \chi_{\mu}^{\vec{L}} (\vec{r}) ~ \chi_{\nu}^{\vec{L}\,'} (\vec{r})\\
m_y(\vec{r}) & = & \sum_{\mu \nu} \sum_{\vec{L} \vec{L}\,'} 2 \left[\textbf{D}_{\text{IS}}^{\alpha \beta} \right]_{\mu \nu}^{\vec{L} \vec{L}\,'} \chi_{\mu}^{\vec{L}} (\vec{r}) ~ \chi_{\nu}^{\vec{L}\,'} (\vec{r})  \\
m_z(\vec{r}) & = & \sum_{\mu \nu} \sum_{\vec{L} \vec{L}\,'} \left[\textbf{D}_{\text{RS}}^{\alpha \alpha} - \textbf{D}_{\text{RS}}^{\beta \beta}  \right]_{\mu \nu}^{\vec{L} \vec{L}\,'} \chi_{\mu}^{\vec{L}} (\vec{r}) ~ \chi_{\nu}^{\vec{L}\,'} (\vec{r}).
\end{eqnarray}
The particle current density $\vec{j}_{\text{p}}$ and the spin-current densities
$\vec{j}_{u}$, with $u \in \{x,y,z \}$, are obtained from the antisymmetric linear combinations
\begin{eqnarray}
\allowdisplaybreaks
\vec{j}_{\text{p}}(\vec{r}) & = & - \frac{\text{i}}{2} \sum_{\mu \nu} \sum_{\vec{L} \vec{L}\,'} \left[\textbf{D}_{\text{IA}}^{\alpha \alpha} + \textbf{D}_{\text{IA}}^{\beta \beta}  \right]_{\mu \nu}^{\vec{L} \vec{L}\,'} \xi_{\mu \nu}^{\vec{L} \vec{L}\,'} \\
\vec{j}_{x}(\vec{r}) & = & - \frac{\text{i}}{2} \sum_{\mu \nu} \sum_{\vec{L} \vec{L}\,'} 2 \left[\textbf{D}_{\text{IA}}^{\alpha \beta} \right]_{\mu \nu}^{\vec{L} \vec{L}\,'} \xi_{\mu \nu}^{\vec{L} \vec{L}\,'} \\
\vec{j}_y(\vec{r}) & = &  - \frac{\text{i}}{2} \sum_{\mu \nu} \sum_{\vec{L} \vec{L}\,'} 2 \left[\textbf{D}_{\text{RA}}^{\alpha \beta} \right]_{\mu \nu}^{\vec{L} \vec{L}\,'}
\xi_{\mu \nu}^{\vec{L} \vec{L}\,'} \\
\vec{j}_z(\vec{r}) & = &  - \frac{\text{i}}{2} \sum_{\mu \nu} \sum_{\vec{L} \vec{L}\,'} \left[\textbf{D}_{\text{IA}}^{\alpha \alpha} - \textbf{D}_{\text{IA}}^{\beta \beta}  \right]_{\mu \nu}^{\vec{L} \vec{L}\,'} \xi_{\mu \nu}^{\vec{L} \vec{L}\,'}
\end{eqnarray}
with 
\begin{equation}
\xi_{\mu \nu}^{\vec{L} \vec{L}\,'} =  \left\{ \left[\vec{\nabla} \chi_{\mu}^{\vec{L}} (\vec{r}) \right] ~ \chi_{\nu}^{\vec{L}\,'} (\vec{r}) - \chi_{\mu}^{\vec{L}} (\vec{r}) ~ \left[ \vec{\nabla}  \chi_{\nu}^{\vec{L}\,'} (\vec{r}) \right] \right\}.
\end{equation}

Following our molecular ansatz, \cite{Holzer.Franke.ea:Current.2022}
the scalar part of the XC potential is obtained as
\begin{equation}
\begin{split}
\label{eq:Vxc0}
& V^{\text{XC}, \vec{L} \vec{L}\,'}_{\mu \nu, 0} =  \frac{1}{2} \int \left[
\frac{\partial g^{\text{XC}}}{\partial n_{\uparrow}} +
\frac{\partial g^{\text{XC}}}{\partial n_{\downarrow}}
\right] \chi_{\mu}^{\vec{L}} (\vec{r}) ~ \chi_{\nu}^{\vec{L}\,'} (\vec{r}) ~ \textrm{d}^3 r \\
& + \frac{1}{2} \int \left[
\frac{|\vec{j}_{{\uparrow}}|^{~2}}{2 n_{\uparrow}^2} 
\frac{\partial g^{\text{XC}}}{\partial n_{\downarrow}} +
\frac{|\vec{j}_{{\downarrow}}|^{~2}}{2 n_{\downarrow}^2} 
\frac{\partial g^{\text{XC}}}{\partial \tilde{\tau}_{\downarrow}}
\right] \chi_{\mu}^{\vec{L}} (\vec{r}) ~ \chi_{\nu}^{\vec{L}\,'} (\vec{r}) ~ \textrm{d}^3 r \\
& - \frac{1}{2} \int
\left[ 
2\frac{\partial g^{\text{XC}}}{\partial \gamma_{\uparrow \uparrow}} \vec{\nabla} n_{\uparrow} +
2\frac{\partial g^{\text{XC}}}{\partial \gamma_{\downarrow \downarrow}} \vec{\nabla} n_{\downarrow}  +
\frac{\partial g^{\text{XC}}}{\partial \gamma_{\uparrow \downarrow}} (\vec{\nabla} n_{\uparrow} + \vec{\nabla} n_{\downarrow}) \right] \\
 & \left[ \left\{ \vec{\nabla} \chi_{\mu}^{\vec{L}} (\vec{r}) \right\} \chi_{\nu}^{\vec{L}\,'} (\vec{r})
+ \chi_{\mu}^{\vec{L}} (\vec{r}) \left\{ \vec{\nabla} \chi_{\nu}^{\vec{L}\,'} (\vec{r}) \right\}
\right]  \textrm{d}^3 r \\
& + \int \frac{1}{2} \left[ 
\frac{\partial g^{\text{XC}}}{\partial \tilde{\tau}_{\uparrow}} + 
\frac{\partial g^{\text{XC}}}{\partial \tilde{\tau}_{\downarrow}}
\right]
\left[ \vec{\nabla} \chi_{\mu}^{\vec{L}} (\vec{r})  \right] \cdot
\left[ \vec{\nabla} \chi_{\nu}^{\vec{L}\,'} (\vec{r})  \right]  \textrm{d}^3 r \\
& + \int \frac{\text{i}}{2} \left[
\frac{\vec{j}_{{\uparrow}}}{n_{\uparrow}}
\frac{\partial g^{\text{XC}}}{\partial \tilde{\tau}_{\uparrow}} + 
\frac{\vec{j}_{{\downarrow}}}{n_{\downarrow}}
\frac{\partial g^{\text{XC}}}{\partial \tilde{\tau}_{\downarrow}} \right]
\xi_{\mu \nu}^{\vec{L} \vec{L}\,'} \textrm{d}^3 r
\end{split}
\end{equation}
and the spin-magnetization part with $u \in \{x, y, z\}$ reads
\begin{equation}
\label{eq:Vxcm}
\begin{split}
& V^{\text{XC},  \vec{L} \vec{L}\,'}_{\mu \nu, u} = \frac{m_u}{2 s}  \int \left[
\frac{\partial g^{\text{XC}}}{\partial n_{\uparrow}} -
\frac{\partial g^{\text{XC}}}{\partial n_{\downarrow}}
\right] \chi_{\mu}^{\vec{L}}  (\vec{r}) ~ \chi_{\nu}^{\vec{L}\,'} (\vec{r}) ~ \textrm{d}^3 r \\
& + \frac{m_u}{2 s}  \int \left[
\frac{|\vec{j}_{{\uparrow}}|^{~2}}{2 n_{\uparrow}^2} 
\frac{\partial g^{\text{XC}}}{\partial n_{\downarrow}} -
\frac{|\vec{j}_{{\downarrow}}|^{~2}}{2 n_{\downarrow}^2} 
\frac{\partial g^{\text{XC}}}{\partial \tilde{\tau}_{\downarrow}}
\right] \chi_{\mu}^{\vec{L}}  (\vec{r}) ~ \chi_{\nu}^{\vec{L}\,'} (\vec{r}) ~ \textrm{d}^3 r \\
& -  \frac{m_u}{2 s} \int
\left[ 
2\frac{\partial g^{\text{XC}}}{\partial \gamma_{\uparrow \uparrow}} \vec{\nabla} n_{\uparrow} -
2\frac{\partial g^{\text{XC}}}{\partial \gamma_{\downarrow \downarrow}} \vec{\nabla} n_{\downarrow}  -
\frac{\partial g^{\text{XC}}}{\partial \gamma_{\uparrow \downarrow}} (\vec{\nabla} n_{\uparrow} - \vec{\nabla} n_{\downarrow}) \right] \\
 & \left[ \left\{ \vec{\nabla} \chi_{\mu}^{\vec{L}}  (\vec{r}) \right\} \chi_{\nu}^{\vec{L}\,'} (\vec{r})
+ \chi_{\mu}^{\vec{L}}  (\vec{r}) \left\{ \vec{\nabla} \chi_{\nu}^{\vec{L}\,'} (\vec{r}) \right\}
\right]  \textrm{d}^3 r \\
& + \int \frac{m_u}{2 s} \left[ 
\frac{\partial g^{\text{XC}}}{\partial \tilde{\tau}_{\uparrow}} - 
\frac{\partial g^{\text{XC}}}{\partial \tilde{\tau}_{\downarrow}}
\right]
\left[ \vec{\nabla} \chi_{\mu}^{\vec{L}}  (\vec{r})  \right] \cdot
\left[ \vec{\nabla} \chi_{\nu}^{\vec{L}\,'} (\vec{r})  \right]  \textrm{d}^3 r \\
& +  \text{i} \int \frac{m_u}{2 s} \left[
\frac{\vec{j}_{{\uparrow}}}{n_{\uparrow}}
\frac{\partial g^{\text{XC}}}{\partial \tilde{\tau}_{\uparrow}} - 
\frac{\vec{j}_{{\downarrow}}}{n_{\downarrow}}
\frac{\partial g^{\text{XC}}}{\partial \tilde{\tau}_{\downarrow}} \right]
\xi_{\mu \nu}^{\vec{L} \vec{L}\,'}  \textrm{d}^3 r \textrm{.}
\end{split}
\end{equation}

\begin{figure*}[t]
\includegraphics[width=1.0\linewidth]{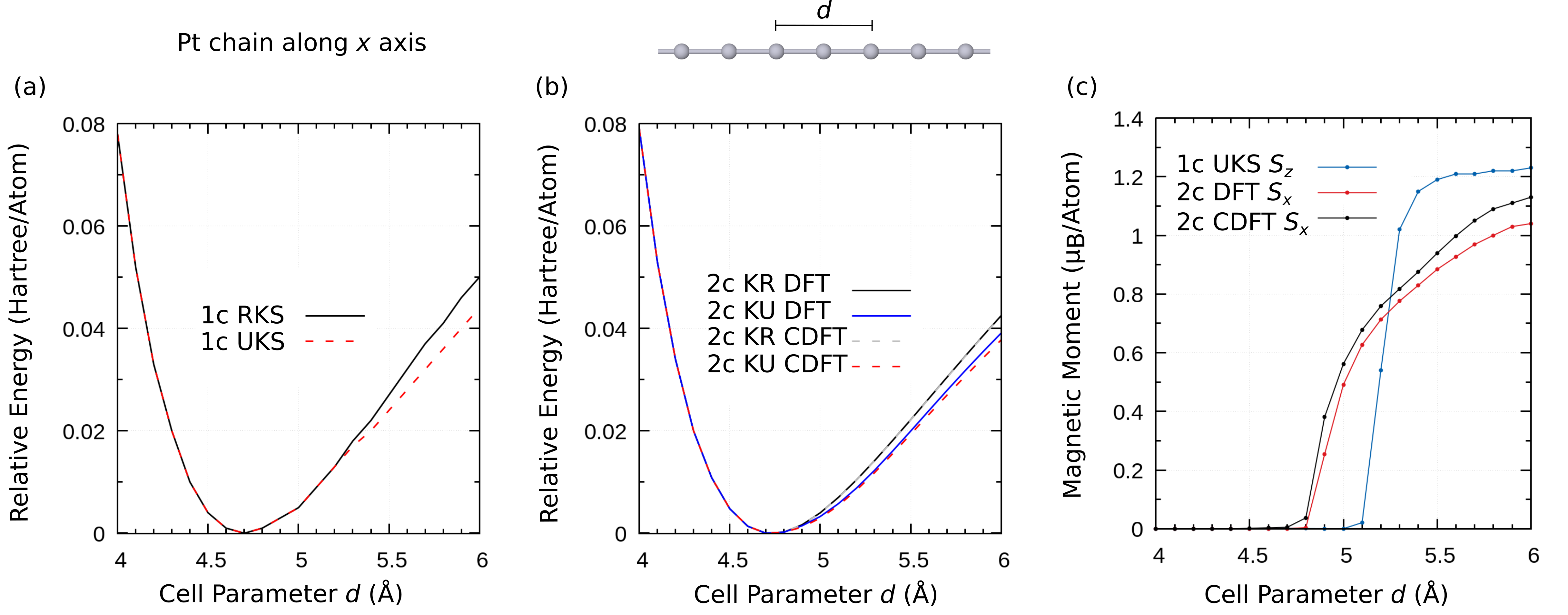}
\caption{
(a) Dependence of the energy on the cell parameter in units of Hartree per atom for
one-component (1c) restricted Kohn--Sham (RKS) and 1c unrestriced Kohn--Sham (UKS)
at the TPSS/dhf-SVP-2c level.
(b) Dependence of the energy on the cell parameter in units of Hartree per atom for
2c KR and 2c KU with the spin aligned along $x$ ($S_x$).
(c) Magnetic moment in units of Bohr's magneton $\mu_{\textrm{B}}$ per atom for the
spin contribution of 1c UKS, 2c KU $S_x$ DFT and CDFT approach.
Periodicity is along the $x$ direction for one-dimensional systems. \cite{TURBOMOLE-manual}
The open-shell solutions are energetically favored compared to the respective
closed-shell solutions and the $S_x$ alignment is preferred over $S_{y,z}$.
Results without the current-independent TPSS functional are taken from
Ref.~\citenum{Franzke.Schosser.ea:Efficient.2024}. Detailed results
are listed in the Supplementary Material.
}
\label{fig:pt}
\end{figure*}

The spin blocks of the Kohn--Sham equations are formed by combining
the scalar and spin-magnetization contributions with the respective Pauli
matrices, c.f.~Refs.\citenum{Holzer.Franke.ea:Current.2022} and
\citenum{Franzke.Schosser.ea:Efficient.2024}. After transformation to
the $k$ space, the Kohn--Sham equations can be solved as usual.

For non-magnetic or closed-shell system, $\vec{m}$ and $\vec{j}_{\text{p}}$ vanish
so that a Kramers-restricted framework can be applied and time-reversal
symmetry \cite{Kasper.Jenkins.ea:Perspective.2020} may be exploited.
However, the spin current densities are still non-zero and hence the
spin-up and spin-down quantities $\vec{j}_{\uparrow, \downarrow}$
contribute to the XC potential through the diamagnetic or quadratic
terms. \cite{Holzer.Franke.ea:Current.2022}

The CDFT approach outlined herein is implemented in the riper module
\cite{Burow.Sierka.ea:Resolution.2009, Burow.Sierka:Linear.2011,
Lazarski.Burow.ea:Density.2015, Lazarski.Burow.ea:Density.2016,
Grajciar:Low.2015, Becker.Sierka:Density.2019, Irmler.Burow.ea:Robust.2018, 
Franzke.Schosser.ea:Efficient.2024} of TURBOMOLE. \cite{Ahlrichs.Bar.ea:Electronic.1989, Franzke.Holzer.ea:TURBOMOLE.2023, TURBOMOLE} The numerical integration
of the exchange-correlation potential is carried with the algorithm
of Ref.~\citenum{Burow.Sierka:Linear.2011}. Note that the current
density also leads to antisymmetric contributions. Integration
weights are constructed according to Stratmann \textit{et al.}
\cite{Stratmann.Scuseria.ea:Achieving.1996}
Geometry gradients and stress tensors are implemented based
on previous work, as it essentially involves further derivatives
of the basis functions. \cite{Lazarski.Burow.ea:Density.2016,
Becker.Sierka:Density.2019, Franzke.Schosser.ea:Efficient.2024}
We note that all integrals and the exchange potential for the
Kohn--Sham equations are evaluated in the position space, which
allows to exploit sparsity. The increase in the computational
cost by calculating the current density contributions on a grid
is modest---especially compared to the inclusion of the current
density through Hartree--Fock exchange with hybrid functions.

\section{Computational Methods}
\label{sec:computational-methods}
First, we study the impact of the current density on the magnetic
transition of one-dimensional linear Pt chains. \cite{Delin.Tosatti:Magnetic.2003,
Delin.Tosatti:Emerging.2004, Fernandez-Rossier.Jacob.ea:Transport.2005,
Smogunov.Dal-Corso.ea:Colossal.2008, Garca-Suarez.Manrique.ea:Anisotropic.2009}
Two Pt atoms are placed in the unit cell with the cell parameter $d$ ranging
from 4.0\,\AA \ to 6.0\,\AA.
Calculations are performed at the TPSS/dhf-SVP-2c \cite{Tao.Perdew.ea:Climbing.2003, 
Weigend.Baldes:Segmented.2010} level employing Dirac--Fock effective
core potentials (DF-ECPs), \cite{Figgen.Peterson.ea:Energy-consistent.2009}
replacing 60 electrons (ECP-60). A Gaussian smearing of 0.01\,Hartree \cite{Kresse.Furthmuller:Efficiency.1996} is used and a $k$-mesh with 32 points
is applied. Integration grids, convergence settings, etc.\  are given in
the Supplementary Material. 
We note in passing that the Karlsruhe dhf-type basis sets were
optimized for discrete systems, \cite{Weigend.Baldes:Segmented.2010}
however, the corresponding pob-type basis sets for periodic calculations
\cite{Peintinger.Oliveira.ea:Consistent.2013, Laun.Vilela-Oliveira.ea:Consistent.2018,
Vilela-Oliveira.Laun.ea:BSSE-correction.2019, Laun.Bredow:BSSE-corrected.2021,
Laun.Bredow:BSSE-corrected.2022, Seidler.Laun.ea:BSSE-corrected.2023}
are not yet available with tailored extensions for spin--orbit
two-component calculations. \cite{Armbruster.Klopper.ea:Basis-set.2006}
Therefore and for consistency with previous studies, we apply
the dhf-type basis sets.

Second, the band gaps and the Rashba splitting of the transition-metal dichalcogenide
monolayers MoCh$_2$ and WCh$_2$ (Ch = S, Se, Te) in the hexagonal (2H) phase
are studied with the M06-L, \cite{Zhao.Truhlar:new.2006}
r$^2$SCAN, \cite{Furness.Kaplan.ea:Accurate.2020, Furness.Kaplan.ea:Correction.2020}
TASK, \cite{Aschebrock.Kummel:Ultranonlocality.2019}
TPSS, \cite{Tao.Perdew.ea:Climbing.2003}
Tao--Mo, \cite{Tao.Mo:Accurate.2016}
and PKZB \cite{Perdew.Kurth.ea:Accurate.1999}
functionals. The GGA PBE \cite{Perdew.Burke.ea:Generalized.1996} 
serves as reference. The dhf-TZVP-2c basis set \cite{Weigend.Baldes:Segmented.2010}
is applied with DF-ECPs for Mo (ECP-28), W (ECP-60), Se (ECP-10), Te (ECP-28).
\cite{Peterson.Figgen.ea:Systematically.2003, Peterson.Figgen.ea:Energy-consistent.2007, 
Figgen.Peterson.ea:Energy-consistent.2009}
Structures are taken from Ref.~\citenum{Miro.Audiffred.ea:atlas.2014}.
A $k$-mesh of $33 \crossproduct 33$ points is applied.

Third, the CDFT framework is applied to complement our previous meta-GGA
study on silver halides. \cite{Franzke.Schosser.ea:Efficient.2024}
Here, we consider the TPSS, \cite{Tao.Perdew.ea:Climbing.2003}
revTPSS, \cite{Perdew.Ruzsinszky.ea:Workhorse.2009, Perdew.Ruzsinszky.ea:Erratum.2011}
Tao--Mo, \cite{Tao.Mo:Accurate.2016}
PKZB, \cite{Perdew.Kurth.ea:Accurate.1999}
r$^2$SCAN, \cite{Furness.Kaplan.ea:Accurate.2020, Furness.Kaplan.ea:Correction.2020}
and M06-L. \cite{Zhao.Truhlar:new.2006}
We use the dhf-SVP basis set \cite{Weigend.Baldes:Segmented.2010} 
and DF-ECPs are applied for Ag (ECP-28) and I (ECP-28). 
\cite{Figgen.Rauhut.ea:Energy-consistent.2005, Peterson.Shepler.ea:On.2006}
A $k$-mesh of $7 \times 7 \times 7$ is employed.

\section{Results and Discussion}
\label{sec:results}

\subsection{Linear Pt Chain}
\label{subsec:pt-chain}

The one-dimensional linear Pt chain is a common reference system for
a transition to a magnetic system. The closed-shell configuration 
constitutes the electronic ground state with a small cell parameter,
whereas the magnetic open-shell solution becomes lower in energy with
increasing cell size. \cite{Delin.Tosatti:Magnetic.2003,
Delin.Tosatti:Emerging.2004,
Fernandez-Rossier.Jacob.ea:Transport.2005, 
Smogunov.Dal-Corso.ea:Colossal.2008,
Garca-Suarez.Manrique.ea:Anisotropic.2009} This is also confirmed
at the scalar and spin--orbit TPSS level in Fig.~\ref{fig:pt}.
For small cell parameters from $d = 4.0 $ to $d = 4.9 $\,\AA,
the open-shell initial guess converges to a closed-shell
non-magnetic solution in the self-consistent field (SCF)
procedure. The most favorable total energy is found for
$d = 4.7 $\,\AA\ in agreement with previous studies based on
GGA functionals. \cite{Smogunov.Dal-Corso.ea:Colossal.2008, 
Franzke.Schosser.ea:Efficient.2024} Further, CDFT and DFT
lead to a very similar potential energy surface or a
similar behavior of the relative energy with respect to
the cell parameter $d$.

At the scalar level, the transition to a magnetic material
occurs between $d \approx 5.2$\,\AA\ and $d \approx 5.3$\,\AA.
Inclusion of spin--orbit coupling shifts this transition to a
smaller $d$ parameter of about $5.0$\,\AA.
Here, the current-dependent variant of TPSS (cTPSS) leads to
a lower total energy for both closed-shell and open-shell
solutions, i.e.\ a more negative energy, and a
larger magnetic moment. The impact of the current density is
generally larger for the magnetic solution than for the
closed-shell state, see also the Supplementary Material for
detailed results. For the closed-shell solution, the
difference in the total energy by inclusion of the current
density is too small to be visible in panel (b) of
Fig.~\ref{fig:pt}. This finding is qualitatively in line
with our previous studies on molecular systems.
\cite{Holzer.Franke.ea:Current.2022}

Inclusion of the current density also consistently
leads to a larger magnetic moment. For instance, 
TPSS predicts a magnetic moment of 1.04\,\textmu$_{\text{B}}$/atom
at $d = 6.0$\,\AA, whereas a value of 1.13\,\textmu$_{\text{B}}$/atom
is found with cTPSS. Generally, an increase in the range of
5--8\,\% is observed after the transition to a magnetic solution.

\subsection{Transition-Metal Dichalcogenide Monolayers}
\label{susec:tmdc}

Transition-metal dichalcogenide monolayers have
many interesting physical properties such as the
quantum spin Hall, \cite{Liu.Feng.ea:Quantum.2011}
non-linear anomalous Hall, \cite{Kang.Li.ea:Nonlinear.2019}
and Rashba effects. \cite{Zhu.Cheng.ea:Giant.2011}
In the H2 phase, time-reversal symmetry holds for
the non-magnetic systems but space-inversion symmetry
is lost. Thus, spin--orbit coupling lifts the degeneracy
of the valence band at the K point in the Brillouin zone.
For the Mo and W systems, this Rashba splitting is very
pronounced and values between 0.1 and 0.5\,eV are
obtained with relativistic all-electron methods.
\cite{Miro.Audiffred.ea:atlas.2014, Kadek.Wang.ea:Band.2023}

\begin{table}[t]
\caption{Band gaps and Rashba splittings of the valence band
at the K point for transition-metal dichalcogenide monolayers
at the 1c DFT, 2c DFT, and 2c CDFT level with the dhf-TZVP-2c
basis set and DF-ECPs for all atoms. All values in eV.
Results for MoS$_2$ and WS$_2$ are listed in the Supplementary
Material.}
\label{tab:tmdc}
\begin{tabular}{@{\extracolsep{2pt}}ll
S[table-format = 1.3]
S[table-format = 1.3]
S[table-format = 1.3]
S[table-format = 1.3]
S[table-format = 1.3]
@{}}
\toprule
 & & \multicolumn{3}{S}{\text{Band Gap}} & \multicolumn{2}{S}{\text{Rashba Splitting}} \\
 \cmidrule{3-5} \cmidrule{6-7}
 {\text{System}} & {\text{DFA}} & {\text{1c DFT}} & {\text{2c DFT}} & {\text{2c CDFT}} & {\text{2c DFT}} & {\text{2c CDFT}} \\
\midrule
MoSe$_2$ & PBE & 1.554 & 1.453 & {\text{--}} & 0.166 & {\text{--}} \\
 & PKZB & 1.558 & 1.455 & 1.455 & 0.166 & 0.169 \\
 & Tao--Mo & 1.561 & 1.459 & 1.458 & 0.166 & 0.170 \\
 & TPSS & 1.563 & 1.459 & 1.458 & 0.167 & 0.174 \\
 & M06-L & 1.563 & 1.460 & 1.455 & 0.170 & 0.181 \\
 & r$^2$SCAN & 1.657 & 1.554 & 1.544 & 0.166 & 0.189 \\
 & TASK & 1.735 & 1.629 & 1.607 & 0.172 & 0.218 \\
MoTe$_2$ & PBE & 1.157 & 1.039 & {\text{--}} & 0.185 & {\text{--}} \\
 & PKZB & 1.160 & 1.042 & 1.042 & 0.186 & 0.189 \\
 & Tao--Mo & 1.168 & 1.051 & 1.049 & 0.185 & 0.188 \\
 & TPSS & 1.170 & 1.051 & 1.048 & 0.185 & 0.195 \\
 & M06-L & 1.182 & 1.060 & 1.053 & 0.192 & 0.209 \\
 & r$^2$SCAN & 1.235 & 1.116 & 1.101 & 0.186 & 0.218 \\
 & TASK & 1.287 & 1.161 & 1.126 & 0.197 & 0.269 \\
WSe$_2$ & PBE & 1.627 & 1.324 & {\text{--}} & 0.409 & {\text{--}} \\
 & PKZB & 1.652 & 1.377 & 1.374 & 0.407 & 0.410 \\
 & Tao--Mo & 1.667 & 1.371 & 1.367 & 0.403 & 0.411 \\
 & TPSS & 1.660 & 1.379 & 1.371 & 0.409 & 0.420 \\
 & M06-L & 1.668 & 1.413 & 1.405 & 0.422 & 0.440 \\
 & r$^2$SCAN & 1.774 & 1.482 & 1.452 & 0.419 & 0.457 \\
 & TASK & 1.867 & 1.604 & 1.571 & 0.440 & 0.514 \\
WTe$_2$ & PBE & 1.219 & 0.933 & {\text{--}} & 0.423 & {\text{--}} \\
 & PKZB & 1.227 & 0.953 & 0.951 & 0.423 & 0.427 \\
 & Tao--Mo & 1.241 & 0.963 & 0.957 & 0.419 & 0.426 \\
 & TPSS & 1.237 & 0.960 & 0.954 & 0.422 & 0.435 \\
 & M06-L & 1.253 & 0.968 & 0.957 & 0.431 & 0.456 \\
 & r$^2$SCAN & 1.325 & 1.039 & 1.005 & 0.436 & 0.483 \\
 & TASK & 1.388 & 1.096 & 1.050 & 0.465 & 0.566 \\
\bottomrule
\end{tabular}
\end{table}

Band gaps and Rashba splittings at the K point obtained
with DFT and CDFT approaches are listed in Tab.~\ref{tab:tmdc}.
One the one hand, the impact on the band gaps is rather small,
with TASK and r$^2$SCAN showing the largest changes but not
exceeding 0.1\,eV. Here, the deviation between the different
DFAs is far larger. One the other hand, the Rashba splitting
is very sensitive to the inclusion of the current density.
For instance, the results change from 0.436\,eV to 0.483\,eV
and 0.465\,eV to 0.566\,eV for WTe$_2$ with r$^2$SCAN and
TASK, respectively.

The most pronounced current-density effects are found for TASK
throughout all systems, which is in line with molecular studies. 
\cite{Holzer.Franke.ea:Current.2022, Bruder.Franzke.ea:Zero-Field.2023, 
Franzke.Holzer:Impact.2022} Here, the current density increase
the Rashba splitting by 25\,\% on average. r$^2$SCAN ranks
second in this regard with 13\,\% followed by M06-L with 6\,\%.
For PKZB, Tao--Mo and TPSS, changes of only 1--3\,\% are observed.
Among the different monolayers, MoTe$_2$ leads to the largest
impact of the current density for all DFAs, with a relative
deviation of 36\,\% between DFT and CDFT for TASK.
The impact of the current density for the DFAs can be rationalized
by the enhancement factor. \cite{Franzke.Holzer:Impact.2022,
Grotjahn.Furche.ea:Importance.2022}

\subsection{Silver Halide Crystals}
\label{susec:agx}

\begin{table}[t]
\begin{center}
\caption{Optimized lattice constant $a$ (in \AA, rocksalt structure)
of three-dimensional AgI and band gaps (in eV) at high symmetry points
of the first Brillouin zone with various density functional
approximations and the dhf-SVP basis sets.
A ``c'' indicates the current-dependent variant of the DFA.
Results for DFAs without inclusion of the current density
are taken from Ref.~\citenum{Franzke.Schosser.ea:Efficient.2024},
except for M06-L. Calculations are performed without dispersion
correction (no D3) and with the D3 correction using
Becke--Johnson damping (D3-BJ).
}
\label{tab:agi-dft}
\begin{tabular}{@{\extracolsep{2pt}}ll
S[table-format = 1.3]
S[table-format = -1.3]
S[table-format = -1.3]
S[table-format = -1.3]
S[table-format = -1.3]
@{}}
\toprule
DFA & Dispersion & $a$ & {\text{L--L}} & {\text{\textGamma--\textGamma}} & {\text{X--X}} & {\text{L--X}} \\
\midrule
TPSS        & no D3  & 6.153 & 3.249 & 2.047 & 2.937 & 0.583 \\
cTPSS       & no D3  & 6.150 & 3.244 & 2.052 & 2.940 & 0.579 \\
revTPSS     & no D3  & 6.116 &3.149 & 2.132 & 2.998 & 0.537 \\
crevTPSS    & no D3  & 6.098 & 3.122 & 2.161 & 3.014 & 0.512 \\
Tao--Mo     & no D3  & 6.071 & 3.099 & 2.343 & 3.108 & 0.616 \\
cTao--Mo    & no D3  & 6.069 & 3.086 & 2.339 & 3.109 & 0.619 \\
PKZB        & no D3  & 6.200 & 3.409 & 2.153 & 2.987 & 0.826 \\
cPKZB       & no D3  & 6.210 & 3.422 & 2.138 & 2.979 & 0.839 \\
r$^2$SCAN   & no D3  & 6.159 & 3.776 & 2.501 & 3.237 & 0.911 \\
cr$^2$SCAN  & no D3  & 6.156 & 3.772 & 2.504 & 3.240 & 0.907 \\
M06-L       & no D3  & 6.325 & 3.511 & 1.891 & 2.976 & 1.116 \\
cM06-L      & no D3  & 6.327 & 3.512 & 1.889 & 2.974 & 1.117 \\
TPSS        & D3-BJ  & 5.982 & 2.988 & 2.352 & 3.098 & 0.343 \\
cTPSS       & D3-BJ  & 5.980 & 2.985 & 2.354 & 3.099 & 0.341 \\
revTPSS     & D3-BJ  & 5.949 & 2.883 & 2.443 & 3.156 & 0.291 \\
crevTPSS    & D3-BJ  & 5.952 & 2.889 & 2.435 & 3.152 & 0.296 \\
Tao--Mo     & D3-BJ  & 6.068 & 3.095 & 2.347 & 3.110 & 0.613 \\
cTao--Mo    & D3-BJ  & 6.062 & 3.087 & 2.358 & 3.116 & 0.604 \\
r$^2$SCAN   & D3-BJ  & 6.156 & 3.773 & 2.506 & 3.240 & 0.907 \\
cr$^2$SCAN  & D3-BJ  & 6.155 & 3.771 & 2.506 & 3.241 & 0.905 \\
\bottomrule
\end{tabular}
\end{center}
\end{table}

The band gaps and optimized lattice constant of AgI with various meta-GGAs
are listed in Tab.~\ref{tab:agi-dft}. Results for AgCl and AgBr are
presented in the Supplementary Material.
Overall, the current density is of minor importance for the band
gaps. Changes are in the order of meV. These results are in
qualitative agreement with the two-dimensional MoTe$_2$ monolayer,
which consists of atoms from the same row of the periodic table
of elements.

Likewise the current density does not lead to major changes for
the cell structure and the lattice constant. The small impact on
the lattice constant can be rationalized by the comparably small
impact of spin--orbit coupling on the cell structure. 
\cite{Franzke.Schosser.ea:Efficient.2024}
D3 dispersion correction with Becke--Johnson damping 
\cite{Grimme.Antony.ea:consistent.2010, Grimme.Ehrlich.ea:Effect.2011, 
Goerigk.Hansen.ea:look.2017, Patra.Jana.ea:Efficient.2020,
Ehlert.Huniar.ea:r2SCAN-D4.2021} (D3-BJ) leads to much
larger changes than the application of CDFT.
Therefore, other computational parameters than the
inclusion of the current density for meta-GGAs are
more important for the cell structures of the silver
halide crystals.

\section{Conclusion}
\label{sec:conclusion}
We have extended our previous formulation of spin--orbit
current density functional theory to periodic systems of
arbitrary dimension. The impact of the current density
was assessed for various properties of non-magnetic and
magnetic systems. Here, the band gaps and lattice constants
are not notably affected. In contrast, the Rashba splitting
which is only due to spin--orbit coupling is substantially
affected. The inclusion of the current density for a given
functional may lead to larger changes than the deviation
of the results among different functionals. 

With the present work, CDFT is now applicable to a wide range of
chemical and physical properties of discrete and periodic
systems, including analytic first-order property calculations.
We generally recommend to include the current density for
r$^2$SCAN and the Minnesota functionals, if available in the
used electronic structure code. For TASK, inclusion of the
current density is clearly mandatory as it leads to substantial
changes of the results.

%%%%%%%%%%%%%%%%%%%%%%%%%%%%%%%%%%%%%%%%%%%%%%%%%%%%%%%%%%%%%%%%%%%%%
%% Supplementary material
%%%%%%%%%%%%%%%%%%%%%%%%%%%%%%%%%%%%%%%%%%%%%%%%%%%%%%%%%%%%%%%%%%%%%
\section*{Supplementary Material}
Supporting Information is available with all computational
details and data.

%%%%%%%%%%%%%%%%%%%%%%%%%%%%%%%%%%%%%%%%%%%%%%%%%%%%%%%%%%%%%%%%%%%%%
%% Acknowledgments
%%%%%%%%%%%%%%%%%%%%%%%%%%%%%%%%%%%%%%%%%%%%%%%%%%%%%%%%%%%%%%%%%%%%%
\begin{acknowledgments}
We thank Fabian Pauly (Augsburg) and Marek Sierka (Jena) for
helpful comments.
Y.J.F.\ gratefully acknowledges support via the Walter--Benjamin
programme funded by the Deutsche Forschungsgemeinschaft (DFG, German
Research Foundation) --- 518707327.
C.H.\ gratefully acknowledges funding by the Volkswagen Foundation.
\end{acknowledgments}

%\vspace{0.1cm}

%%%%%%%%%%%%%%%%%%%%%%%%%%%%%%%%%%%%%%%%%%%%%%%%%%%%%%%%%%%%%%%%%%%%%
%% Author Declarations
%%%%%%%%%%%%%%%%%%%%%%%%%%%%%%%%%%%%%%%%%%%%%%%%%%%%%%%%%%%%%%%%%%%%%
\section*{Author Declarations}
\subsection*{Conflict of Interest}
The authors have no conflicts to disclose.

\subsection*{Author Contributions}
\noindent \textbf{Yannick J. Franzke}: Conceptualization (equal); Data curation (lead);
Formal analysis (lead); Investigation (equal); Methodology (lead);
Software (lead); Validation (equal); Visualization (lead);
Writing – original draft (lead); Writing – review \& editing (equal).
\\
\noindent \textbf{Christof Holzer}: Conceptualization (equal); Data curation (supporting);
Formal analysis (supporting); Investigation (equal); Methodology (supporting);
Software (supporting); Validation (equal);
Writing – original draft (supporting); Writing – review \& editing (equal).

%%%%%%%%%%%%%%%%%%%%%%%%%%%%%%%%%%%%%%%%%%%%%%%%%%%%%%%%%%%%%%%%%%%%%
%% Data Availability Statement
%%%%%%%%%%%%%%%%%%%%%%%%%%%%%%%%%%%%%%%%%%%%%%%%%%%%%%%%%%%%%%%%%%%%%
\section*{Data Availability Statement}
The data that support the findings of this study are available within the
article and its supplementary material.

%%%%%%%%%%%%%%%%%%%%%%%%%%%%%%%%%%%%%%%%%%%%%%%%%%%%%%%%%%%%%%%%%%%%%
%% References
%%%%%%%%%%%%%%%%%%%%%%%%%%%%%%%%%%%%%%%%%%%%%%%%%%%%%%%%%%%%%%%%%%%%%
\section*{References}
\bibliography{literature}% Produces the bibliography via BibTeX.

\begin{thebibliography}{107}%
\makeatletter
\providecommand \@ifxundefined [1]{%
 \@ifx{#1\undefined}
}%
\providecommand \@ifnum [1]{%
 \ifnum #1\expandafter \@firstoftwo
 \else \expandafter \@secondoftwo
 \fi
}%
\providecommand \@ifx [1]{%
 \ifx #1\expandafter \@firstoftwo
 \else \expandafter \@secondoftwo
 \fi
}%
\providecommand \natexlab [1]{#1}%
\providecommand \enquote  [1]{``#1''}%
\providecommand \bibnamefont  [1]{#1}%
\providecommand \bibfnamefont [1]{#1}%
\providecommand \citenamefont [1]{#1}%
\providecommand \href@noop [0]{\@secondoftwo}%
\providecommand \href [0]{\begingroup \@sanitize@url \@href}%
\providecommand \@href[1]{\@@startlink{#1}\@@href}%
\providecommand \@@href[1]{\endgroup#1\@@endlink}%
\providecommand \@sanitize@url [0]{\catcode `\\12\catcode `\$12\catcode
  `\&12\catcode `\#12\catcode `\^12\catcode `\_12\catcode `\%12\relax}%
\providecommand \@@startlink[1]{}%
\providecommand \@@endlink[0]{}%
\providecommand \url  [0]{\begingroup\@sanitize@url \@url }%
\providecommand \@url [1]{\endgroup\@href {#1}{\urlprefix }}%
\providecommand \urlprefix  [0]{URL }%
\providecommand \Eprint [0]{\href }%
\providecommand \doibase [0]{http://dx.doi.org/}%
\providecommand \selectlanguage [0]{\@gobble}%
\providecommand \bibinfo  [0]{\@secondoftwo}%
\providecommand \bibfield  [0]{\@secondoftwo}%
\providecommand \translation [1]{[#1]}%
\providecommand \BibitemOpen [0]{}%
\providecommand \bibitemStop [0]{}%
\providecommand \bibitemNoStop [0]{.\EOS\space}%
\providecommand \EOS [0]{\spacefactor3000\relax}%
\providecommand \BibitemShut  [1]{\csname bibitem#1\endcsname}%
\let\auto@bib@innerbib\@empty
%</preamble>
\bibitem [{\citenamefont {Burke}(2012)}]{Burke:Perspective.2012}%
  \BibitemOpen
  \bibfield  {author} {\bibinfo {author} {\bibfnamefont {K.}~\bibnamefont
  {Burke}},\ }\href {\doibase 10.1063/1.4704546} {\bibfield  {journal}
  {\bibinfo  {journal} {J. Chem. Phys.}\ }\textbf {\bibinfo {volume} {136}},\
  \bibinfo {pages} {150901} (\bibinfo {year} {2012})}\BibitemShut {NoStop}%
\bibitem [{\citenamefont {Becke}(2014)}]{Becke:Perspective.2014}%
  \BibitemOpen
  \bibfield  {author} {\bibinfo {author} {\bibfnamefont {A.~D.}\ \bibnamefont
  {Becke}},\ }\href {\doibase 10.1063/1.4869598} {\bibfield  {journal}
  {\bibinfo  {journal} {J. Chem. Phys.}\ }\textbf {\bibinfo {volume} {140}},\
  \bibinfo {pages} {18A301} (\bibinfo {year} {2014})}\BibitemShut {NoStop}%
\bibitem [{\citenamefont {Sun}, \citenamefont {Ruzsinszky},\ and\ \citenamefont
  {Perdew}(2015)}]{Sun.Ruzsinszky.ea:Strongly.2015}%
  \BibitemOpen
  \bibfield  {author} {\bibinfo {author} {\bibfnamefont {J.}~\bibnamefont
  {Sun}}, \bibinfo {author} {\bibfnamefont {A.}~\bibnamefont {Ruzsinszky}}, \
  and\ \bibinfo {author} {\bibfnamefont {J.~P.}\ \bibnamefont {Perdew}},\
  }\href {\doibase 10.1103/PhysRevLett.115.036402} {\bibfield  {journal}
  {\bibinfo  {journal} {Phys. Rev. Lett.}\ }\textbf {\bibinfo {volume} {115}},\
  \bibinfo {pages} {036402} (\bibinfo {year} {2015})}\BibitemShut {NoStop}%
\bibitem [{\citenamefont {Mardirossian}\ and\ \citenamefont
  {Head-Gordon}(2017)}]{Mardirossian.Head-Gordon:Thirty.2017}%
  \BibitemOpen
  \bibfield  {author} {\bibinfo {author} {\bibfnamefont {N.}~\bibnamefont
  {Mardirossian}}\ and\ \bibinfo {author} {\bibfnamefont {M.}~\bibnamefont
  {Head-Gordon}},\ }\href {\doibase 10.1080/00268976.2017.1333644} {\bibfield
  {journal} {\bibinfo  {journal} {Mol. Phys.}\ }\textbf {\bibinfo {volume}
  {115}},\ \bibinfo {pages} {2315} (\bibinfo {year} {2017})}\BibitemShut
  {NoStop}%
\bibitem [{\citenamefont {Tao}\ and\ \citenamefont
  {Mo}(2016)}]{Tao.Mo:Accurate.2016}%
  \BibitemOpen
  \bibfield  {author} {\bibinfo {author} {\bibfnamefont {J.}~\bibnamefont
  {Tao}}\ and\ \bibinfo {author} {\bibfnamefont {Y.}~\bibnamefont {Mo}},\
  }\href {\doibase 10.1103/PhysRevLett.117.073001} {\bibfield  {journal}
  {\bibinfo  {journal} {Phys. Rev. Lett.}\ }\textbf {\bibinfo {volume} {117}},\
  \bibinfo {pages} {073001} (\bibinfo {year} {2016})}\BibitemShut {NoStop}%
\bibitem [{\citenamefont {Hao}\ \emph {et~al.}(2013)\citenamefont {Hao},
  \citenamefont {Sun}, \citenamefont {Xiao}, \citenamefont {Ruzsinszky},
  \citenamefont {Csonka}, \citenamefont {Tao}, \citenamefont {Glindmeyer},\
  and\ \citenamefont {Perdew}}]{Hao.Sun.ea:Performance.2013}%
  \BibitemOpen
  \bibfield  {author} {\bibinfo {author} {\bibfnamefont {P.}~\bibnamefont
  {Hao}}, \bibinfo {author} {\bibfnamefont {J.}~\bibnamefont {Sun}}, \bibinfo
  {author} {\bibfnamefont {B.}~\bibnamefont {Xiao}}, \bibinfo {author}
  {\bibfnamefont {A.}~\bibnamefont {Ruzsinszky}}, \bibinfo {author}
  {\bibfnamefont {G.~I.}\ \bibnamefont {Csonka}}, \bibinfo {author}
  {\bibfnamefont {J.}~\bibnamefont {Tao}}, \bibinfo {author} {\bibfnamefont
  {S.}~\bibnamefont {Glindmeyer}}, \ and\ \bibinfo {author} {\bibfnamefont
  {J.~P.}\ \bibnamefont {Perdew}},\ }\href {\doibase 10.1021/ct300868x}
  {\bibfield  {journal} {\bibinfo  {journal} {J. Chem. Theory Comput.}\
  }\textbf {\bibinfo {volume} {9}},\ \bibinfo {pages} {355} (\bibinfo {year}
  {2013})}\BibitemShut {NoStop}%
\bibitem [{\citenamefont {Mo}\ \emph {et~al.}(2017)\citenamefont {Mo},
  \citenamefont {Car}, \citenamefont {Staroverov}, \citenamefont {Scuseria},\
  and\ \citenamefont {Tao}}]{Mo.Car.ea:Assessment.2017}%
  \BibitemOpen
  \bibfield  {author} {\bibinfo {author} {\bibfnamefont {Y.}~\bibnamefont
  {Mo}}, \bibinfo {author} {\bibfnamefont {R.}~\bibnamefont {Car}}, \bibinfo
  {author} {\bibfnamefont {V.~N.}\ \bibnamefont {Staroverov}}, \bibinfo
  {author} {\bibfnamefont {G.~E.}\ \bibnamefont {Scuseria}}, \ and\ \bibinfo
  {author} {\bibfnamefont {J.}~\bibnamefont {Tao}},\ }\href {\doibase
  10.1103/PhysRevB.95.035118} {\bibfield  {journal} {\bibinfo  {journal} {Phys.
  Rev. B}\ }\textbf {\bibinfo {volume} {95}},\ \bibinfo {pages} {035118}
  (\bibinfo {year} {2017})}\BibitemShut {NoStop}%
\bibitem [{\citenamefont {Goerigk}\ \emph {et~al.}(2017)\citenamefont
  {Goerigk}, \citenamefont {Hansen}, \citenamefont {Bauer}, \citenamefont
  {Ehrlich}, \citenamefont {Najibi},\ and\ \citenamefont
  {Grimme}}]{Goerigk.Hansen.ea:look.2017}%
  \BibitemOpen
  \bibfield  {author} {\bibinfo {author} {\bibfnamefont {L.}~\bibnamefont
  {Goerigk}}, \bibinfo {author} {\bibfnamefont {A.}~\bibnamefont {Hansen}},
  \bibinfo {author} {\bibfnamefont {C.}~\bibnamefont {Bauer}}, \bibinfo
  {author} {\bibfnamefont {S.}~\bibnamefont {Ehrlich}}, \bibinfo {author}
  {\bibfnamefont {A.}~\bibnamefont {Najibi}}, \ and\ \bibinfo {author}
  {\bibfnamefont {S.}~\bibnamefont {Grimme}},\ }\href {\doibase
  10.1039/C7CP04913G} {\bibfield  {journal} {\bibinfo  {journal} {Phys. Chem.
  Chem. Phys.}\ }\textbf {\bibinfo {volume} {19}},\ \bibinfo {pages} {32184}
  (\bibinfo {year} {2017})}\BibitemShut {NoStop}%
\bibitem [{\citenamefont {Franzke}\ and\ \citenamefont
  {Yu}(2022{\natexlab{a}})}]{Franzke.Yu:Hyperfine.2022}%
  \BibitemOpen
  \bibfield  {author} {\bibinfo {author} {\bibfnamefont {Y.~J.}\ \bibnamefont
  {Franzke}}\ and\ \bibinfo {author} {\bibfnamefont {J.~M.}\ \bibnamefont
  {Yu}},\ }\href {\doibase 10.1021/acs.jctc.1c01027} {\bibfield  {journal}
  {\bibinfo  {journal} {J. Chem. Theory Comput.}\ }\textbf {\bibinfo {volume}
  {18}},\ \bibinfo {pages} {323} (\bibinfo {year}
  {2022}{\natexlab{a}})}\BibitemShut {NoStop}%
\bibitem [{\citenamefont {Franzke}\ and\ \citenamefont
  {Yu}(2022{\natexlab{b}})}]{Franzke.Yu:Quasi-Relativistic.2022}%
  \BibitemOpen
  \bibfield  {author} {\bibinfo {author} {\bibfnamefont {Y.~J.}\ \bibnamefont
  {Franzke}}\ and\ \bibinfo {author} {\bibfnamefont {J.~M.}\ \bibnamefont
  {Yu}},\ }\href {\doibase 10.1021/acs.jctc.1c01175} {\bibfield  {journal}
  {\bibinfo  {journal} {J. Chem. Theory Comput.}\ }\textbf {\bibinfo {volume}
  {18}},\ \bibinfo {pages} {2246} (\bibinfo {year}
  {2022}{\natexlab{b}})}\BibitemShut {NoStop}%
\bibitem [{\citenamefont {Becke}(2022)}]{Becke:Density-Functional.2022}%
  \BibitemOpen
  \bibfield  {author} {\bibinfo {author} {\bibfnamefont {A.~D.}\ \bibnamefont
  {Becke}},\ }\href {\doibase 10.1063/5.0091198} {\bibfield  {journal}
  {\bibinfo  {journal} {J. Chem. Phys.}\ }\textbf {\bibinfo {volume} {156}},\
  \bibinfo {pages} {214101} (\bibinfo {year} {2022})}\BibitemShut {NoStop}%
\bibitem [{\citenamefont {Borlido}\ \emph {et~al.}(2019)\citenamefont
  {Borlido}, \citenamefont {Aull}, \citenamefont {Huran}, \citenamefont {Tran},
  \citenamefont {Marques},\ and\ \citenamefont
  {Botti}}]{Borlido.Aull.ea:Large-Scale.2019}%
  \BibitemOpen
  \bibfield  {author} {\bibinfo {author} {\bibfnamefont {P.}~\bibnamefont
  {Borlido}}, \bibinfo {author} {\bibfnamefont {T.}~\bibnamefont {Aull}},
  \bibinfo {author} {\bibfnamefont {A.~W.}\ \bibnamefont {Huran}}, \bibinfo
  {author} {\bibfnamefont {F.}~\bibnamefont {Tran}}, \bibinfo {author}
  {\bibfnamefont {M.~A.~L.}\ \bibnamefont {Marques}}, \ and\ \bibinfo {author}
  {\bibfnamefont {S.}~\bibnamefont {Botti}},\ }\href {\doibase
  10.1021/acs.jctc.9b00322} {\bibfield  {journal} {\bibinfo  {journal} {J.
  Chem. Theory Comput.}\ }\textbf {\bibinfo {volume} {15}},\ \bibinfo {pages}
  {5069} (\bibinfo {year} {2019})}\BibitemShut {NoStop}%
\bibitem [{\citenamefont {Holzer}, \citenamefont {Franzke},\ and\ \citenamefont
  {Kehry}(2021)}]{Holzer.Franzke.ea:Assessing.2021}%
  \BibitemOpen
  \bibfield  {author} {\bibinfo {author} {\bibfnamefont {C.}~\bibnamefont
  {Holzer}}, \bibinfo {author} {\bibfnamefont {Y.~J.}\ \bibnamefont {Franzke}},
  \ and\ \bibinfo {author} {\bibfnamefont {M.}~\bibnamefont {Kehry}},\ }\href
  {\doibase 10.1021/acs.jctc.1c00203} {\bibfield  {journal} {\bibinfo
  {journal} {J. Chem. Theory Comput.}\ }\textbf {\bibinfo {volume} {17}},\
  \bibinfo {pages} {2928} (\bibinfo {year} {2021})}\BibitemShut {NoStop}%
\bibitem [{\citenamefont {Holzer}\ and\ \citenamefont
  {Franzke}(2022)}]{Holzer.Franzke:Local.2022}%
  \BibitemOpen
  \bibfield  {author} {\bibinfo {author} {\bibfnamefont {C.}~\bibnamefont
  {Holzer}}\ and\ \bibinfo {author} {\bibfnamefont {Y.~J.}\ \bibnamefont
  {Franzke}},\ }\href {\doibase 10.1063/5.0100439} {\bibfield  {journal}
  {\bibinfo  {journal} {J. Chem. Phys.}\ }\textbf {\bibinfo {volume} {157}},\
  \bibinfo {pages} {034108} (\bibinfo {year} {2022})}\BibitemShut {NoStop}%
\bibitem [{\citenamefont {Lee}\ \emph {et~al.}(2021)\citenamefont {Lee},
  \citenamefont {Feng}, \citenamefont {Cunha}, \citenamefont {Gonthier},
  \citenamefont {Epifanovsky},\ and\ \citenamefont
  {Head-Gordon}}]{Lee.Feng.ea:Approaching.2021}%
  \BibitemOpen
  \bibfield  {author} {\bibinfo {author} {\bibfnamefont {J.}~\bibnamefont
  {Lee}}, \bibinfo {author} {\bibfnamefont {X.}~\bibnamefont {Feng}}, \bibinfo
  {author} {\bibfnamefont {L.~A.}\ \bibnamefont {Cunha}}, \bibinfo {author}
  {\bibfnamefont {J.~F.}\ \bibnamefont {Gonthier}}, \bibinfo {author}
  {\bibfnamefont {E.}~\bibnamefont {Epifanovsky}}, \ and\ \bibinfo {author}
  {\bibfnamefont {M.}~\bibnamefont {Head-Gordon}},\ }\href {\doibase
  10.1063/5.0069177} {\bibfield  {journal} {\bibinfo  {journal} {J. Chem.
  Phys.}\ }\textbf {\bibinfo {volume} {155}},\ \bibinfo {pages} {164102}
  (\bibinfo {year} {2021})}\BibitemShut {NoStop}%
\bibitem [{\citenamefont {Ghosh}\ \emph {et~al.}(2022)\citenamefont {Ghosh},
  \citenamefont {Jana}, \citenamefont {Rauch}, \citenamefont {Tran},
  \citenamefont {Marques}, \citenamefont {Botti}, \citenamefont {Constantin},
  \citenamefont {Niranjan},\ and\ \citenamefont
  {Samal}}]{Ghosh.Jana.ea:Efficient.2022}%
  \BibitemOpen
  \bibfield  {author} {\bibinfo {author} {\bibfnamefont {A.}~\bibnamefont
  {Ghosh}}, \bibinfo {author} {\bibfnamefont {S.}~\bibnamefont {Jana}},
  \bibinfo {author} {\bibfnamefont {T.}~\bibnamefont {Rauch}}, \bibinfo
  {author} {\bibfnamefont {F.}~\bibnamefont {Tran}}, \bibinfo {author}
  {\bibfnamefont {M.~A.~L.}\ \bibnamefont {Marques}}, \bibinfo {author}
  {\bibfnamefont {S.}~\bibnamefont {Botti}}, \bibinfo {author} {\bibfnamefont
  {L.~A.}\ \bibnamefont {Constantin}}, \bibinfo {author} {\bibfnamefont
  {M.~K.}\ \bibnamefont {Niranjan}}, \ and\ \bibinfo {author} {\bibfnamefont
  {P.}~\bibnamefont {Samal}},\ }\href {\doibase 10.1063/5.0111693} {\bibfield
  {journal} {\bibinfo  {journal} {J. Chem. Phys.}\ }\textbf {\bibinfo {volume}
  {157}},\ \bibinfo {pages} {124108} (\bibinfo {year} {2022})}\BibitemShut
  {NoStop}%
\bibitem [{\citenamefont {Kov\'as}\ \emph {et~al.}(2022)\citenamefont
  {Kov\'as}, \citenamefont {Tran}, \citenamefont {Blaha},\ and\ \citenamefont
  {Madsen}}]{Kovacs.Tran.ea:What.2022}%
  \BibitemOpen
  \bibfield  {author} {\bibinfo {author} {\bibfnamefont {P.}~\bibnamefont
  {Kov\'as}}, \bibinfo {author} {\bibfnamefont {F.}~\bibnamefont {Tran}},
  \bibinfo {author} {\bibfnamefont {P.}~\bibnamefont {Blaha}}, \ and\ \bibinfo
  {author} {\bibfnamefont {G.~K.~H.}\ \bibnamefont {Madsen}},\ }\href {\doibase
  10.1063/5.0098787} {\bibfield  {journal} {\bibinfo  {journal} {J. Chem.
  Phys.}\ }\textbf {\bibinfo {volume} {157}},\ \bibinfo {pages} {094110}
  (\bibinfo {year} {2022})}\BibitemShut {NoStop}%
\bibitem [{\citenamefont {Liang}\ \emph {et~al.}(2022)\citenamefont {Liang},
  \citenamefont {Feng}, \citenamefont {Hait},\ and\ \citenamefont
  {Head-Gordon}}]{Liang.Feng.ea:Revisiting.2022}%
  \BibitemOpen
  \bibfield  {author} {\bibinfo {author} {\bibfnamefont {J.}~\bibnamefont
  {Liang}}, \bibinfo {author} {\bibfnamefont {X.}~\bibnamefont {Feng}},
  \bibinfo {author} {\bibfnamefont {D.}~\bibnamefont {Hait}}, \ and\ \bibinfo
  {author} {\bibfnamefont {M.}~\bibnamefont {Head-Gordon}},\ }\href {\doibase
  10.1021/acs.jctc.2c00160} {\bibfield  {journal} {\bibinfo  {journal} {J.
  Chem. Theory Comput.}\ }\textbf {\bibinfo {volume} {18}},\ \bibinfo {pages}
  {3460} (\bibinfo {year} {2022})}\BibitemShut {NoStop}%
\bibitem [{\citenamefont {Franzke}(2023)}]{Franzke:Reducing.2023}%
  \BibitemOpen
  \bibfield  {author} {\bibinfo {author} {\bibfnamefont {Y.~J.}\ \bibnamefont
  {Franzke}},\ }\href {\doibase 10.1021/acs.jctc.2c01248} {\bibfield  {journal}
  {\bibinfo  {journal} {J. Chem. Theory Comput.}\ }\textbf {\bibinfo {volume}
  {19}},\ \bibinfo {pages} {2010} (\bibinfo {year} {2023})}\BibitemShut
  {NoStop}%
\bibitem [{\citenamefont {Kov\'acs}, \citenamefont {Blaha},\ and\ \citenamefont
  {Madsen}(2023)}]{Kovacs.Blaha.ea:Origin.2023}%
  \BibitemOpen
  \bibfield  {author} {\bibinfo {author} {\bibfnamefont {P.}~\bibnamefont
  {Kov\'acs}}, \bibinfo {author} {\bibfnamefont {P.}~\bibnamefont {Blaha}}, \
  and\ \bibinfo {author} {\bibfnamefont {G.~K.~H.}\ \bibnamefont {Madsen}},\
  }\href {\doibase 10.1063/5.0179260} {\bibfield  {journal} {\bibinfo
  {journal} {J. Chem. Phys.}\ }\textbf {\bibinfo {volume} {159}},\ \bibinfo
  {pages} {244118} (\bibinfo {year} {2023})}\BibitemShut {NoStop}%
\bibitem [{\citenamefont {Lebeda}\ \emph {et~al.}(2023)\citenamefont {Lebeda},
  \citenamefont {Aschebrock}, \citenamefont {Sun}, \citenamefont {Leppert},\
  and\ \citenamefont {K\"ummel}}]{Lebeda.Aschebrock.ea:Right.2023}%
  \BibitemOpen
  \bibfield  {author} {\bibinfo {author} {\bibfnamefont {T.}~\bibnamefont
  {Lebeda}}, \bibinfo {author} {\bibfnamefont {T.}~\bibnamefont {Aschebrock}},
  \bibinfo {author} {\bibfnamefont {J.}~\bibnamefont {Sun}}, \bibinfo {author}
  {\bibfnamefont {L.}~\bibnamefont {Leppert}}, \ and\ \bibinfo {author}
  {\bibfnamefont {S.}~\bibnamefont {K\"ummel}},\ }\href {\doibase
  10.1103/PhysRevMaterials.7.093803} {\bibfield  {journal} {\bibinfo  {journal}
  {Phys. Rev. Mater.}\ }\textbf {\bibinfo {volume} {7}},\ \bibinfo {pages}
  {093803} (\bibinfo {year} {2023})}\BibitemShut {NoStop}%
\bibitem [{\citenamefont {Furness}\ \emph {et~al.}(2015)\citenamefont
  {Furness}, \citenamefont {Verbeke}, \citenamefont {Tellgren}, \citenamefont
  {Stopkowicz}, \citenamefont {Ekstr\"om}, \citenamefont {Helgaker},\ and\
  \citenamefont {Teale}}]{Furness.Verbeke.ea:Current.2015}%
  \BibitemOpen
  \bibfield  {author} {\bibinfo {author} {\bibfnamefont {J.~W.}\ \bibnamefont
  {Furness}}, \bibinfo {author} {\bibfnamefont {J.}~\bibnamefont {Verbeke}},
  \bibinfo {author} {\bibfnamefont {E.~I.}\ \bibnamefont {Tellgren}}, \bibinfo
  {author} {\bibfnamefont {S.}~\bibnamefont {Stopkowicz}}, \bibinfo {author}
  {\bibfnamefont {U.}~\bibnamefont {Ekstr\"om}}, \bibinfo {author}
  {\bibfnamefont {T.}~\bibnamefont {Helgaker}}, \ and\ \bibinfo {author}
  {\bibfnamefont {A.~M.}\ \bibnamefont {Teale}},\ }\href {\doibase
  10.1021/acs.jctc.5b00535} {\bibfield  {journal} {\bibinfo  {journal} {J.
  Chem. Theory Comput.}\ }\textbf {\bibinfo {volume} {11}},\ \bibinfo {pages}
  {4169} (\bibinfo {year} {2015})}\BibitemShut {NoStop}%
\bibitem [{\citenamefont {Tellgren}\ \emph {et~al.}(2014)\citenamefont
  {Tellgren}, \citenamefont {Teale}, \citenamefont {Furness}, \citenamefont
  {Lange}, \citenamefont {Ekstr\"om},\ and\ \citenamefont
  {Helgaker}}]{Tellgren.Teale.ea:Non-perturbative.2014}%
  \BibitemOpen
  \bibfield  {author} {\bibinfo {author} {\bibfnamefont {E.~I.}\ \bibnamefont
  {Tellgren}}, \bibinfo {author} {\bibfnamefont {A.~M.}\ \bibnamefont {Teale}},
  \bibinfo {author} {\bibfnamefont {J.~W.}\ \bibnamefont {Furness}}, \bibinfo
  {author} {\bibfnamefont {K.~K.}\ \bibnamefont {Lange}}, \bibinfo {author}
  {\bibfnamefont {U.}~\bibnamefont {Ekstr\"om}}, \ and\ \bibinfo {author}
  {\bibfnamefont {T.}~\bibnamefont {Helgaker}},\ }\href {\doibase
  10.1063/1.4861427} {\bibfield  {journal} {\bibinfo  {journal} {J. Chem.
  Phys.}\ }\textbf {\bibinfo {volume} {140}},\ \bibinfo {pages} {034101}
  (\bibinfo {year} {2014})}\BibitemShut {NoStop}%
\bibitem [{\citenamefont {Irons}, \citenamefont {David},\ and\ \citenamefont
  {Teale}(2021)}]{Irons.David.ea:Optimizing.2021}%
  \BibitemOpen
  \bibfield  {author} {\bibinfo {author} {\bibfnamefont {T.~J.~P.}\
  \bibnamefont {Irons}}, \bibinfo {author} {\bibfnamefont {G.}~\bibnamefont
  {David}}, \ and\ \bibinfo {author} {\bibfnamefont {A.~M.}\ \bibnamefont
  {Teale}},\ }\href {\doibase 10.1021/acs.jctc.0c01297} {\bibfield  {journal}
  {\bibinfo  {journal} {J. Chem. Theory Comput.}\ }\textbf {\bibinfo {volume}
  {17}},\ \bibinfo {pages} {2166} (\bibinfo {year} {2021})}\BibitemShut
  {NoStop}%
\bibitem [{\citenamefont {Pausch}\ and\ \citenamefont
  {Holzer}(2022)}]{Pausch.Holzer:Linear.2022}%
  \BibitemOpen
  \bibfield  {author} {\bibinfo {author} {\bibfnamefont {A.}~\bibnamefont
  {Pausch}}\ and\ \bibinfo {author} {\bibfnamefont {C.}~\bibnamefont
  {Holzer}},\ }\href {\doibase 10.1021/acs.jpclett.2c01082} {\bibfield
  {journal} {\bibinfo  {journal} {J. Phys. Chem. Lett.}\ }\textbf {\bibinfo
  {volume} {13}},\ \bibinfo {pages} {4335} (\bibinfo {year}
  {2022})}\BibitemShut {NoStop}%
\bibitem [{\citenamefont {Saue}(2005)}]{InB-Saue:2005}%
  \BibitemOpen
  \bibfield  {author} {\bibinfo {author} {\bibfnamefont {T.}~\bibnamefont
  {Saue}},\ }\enquote {\bibinfo {title} {Spin-interactions and the
  non-relativistic limit of electrodynamics},}\ in\ \href {\doibase
  https://doi.org/10.1016/S0065-3276(05)48020-X} {\emph {\bibinfo {booktitle}
  {Advances in Quantum Chemistry}}},\ Vol.~\bibinfo {volume} {48},\ \bibinfo
  {editor} {edited by\ \bibinfo {editor} {\bibfnamefont {J.~R.}\ \bibnamefont
  {Sabin}}, \bibinfo {editor} {\bibfnamefont {E.}~\bibnamefont {Br\"andas}}, \
  and\ \bibinfo {editor} {\bibfnamefont {L.~B.}\ \bibnamefont {Oddershede}}}\
  (\bibinfo  {publisher} {Elsevier Academic Press},\ \bibinfo {address} {San
  Diego, CA, USA},\ \bibinfo {year} {2005})\ pp.\ \bibinfo {pages}
  {383--405}\BibitemShut {NoStop}%
\bibitem [{\citenamefont {Saue}(2011)}]{Saue:Relativistic.2011}%
  \BibitemOpen
  \bibfield  {author} {\bibinfo {author} {\bibfnamefont {T.}~\bibnamefont
  {Saue}},\ }\href {\doibase 10.1002/cphc.201100682} {\bibfield  {journal}
  {\bibinfo  {journal} {ChemPhysChem}\ }\textbf {\bibinfo {volume} {12}},\
  \bibinfo {pages} {3077} (\bibinfo {year} {2011})}\BibitemShut {NoStop}%
\bibitem [{\citenamefont {Pyykk{\"o}}(2012)}]{Pyykko:Relativistic.2012}%
  \BibitemOpen
  \bibfield  {author} {\bibinfo {author} {\bibfnamefont {P.}~\bibnamefont
  {Pyykk{\"o}}},\ }\href {\doibase 10.1146/annurev-physchem-032511-143755}
  {\bibfield  {journal} {\bibinfo  {journal} {Annu. Rev. Phys. Chem.}\ }\textbf
  {\bibinfo {volume} {63}},\ \bibinfo {pages} {45} (\bibinfo {year}
  {2012})}\BibitemShut {NoStop}%
\bibitem [{\citenamefont {Dobson}(1993)}]{Dobson:Alternative.1993}%
  \BibitemOpen
  \bibfield  {author} {\bibinfo {author} {\bibfnamefont {J.~F.}\ \bibnamefont
  {Dobson}},\ }\href {\doibase 10.1063/1.464444} {\bibfield  {journal}
  {\bibinfo  {journal} {J. Chem. Phys.}\ }\textbf {\bibinfo {volume} {98}},\
  \bibinfo {pages} {8870} (\bibinfo {year} {1993})}\BibitemShut {NoStop}%
\bibitem [{\citenamefont {Tao}(2005)}]{Tao:Explicit.2005}%
  \BibitemOpen
  \bibfield  {author} {\bibinfo {author} {\bibfnamefont {J.}~\bibnamefont
  {Tao}},\ }\href {\doibase 10.1103/PhysRevB.71.205107} {\bibfield  {journal}
  {\bibinfo  {journal} {Phys. Rev. B}\ }\textbf {\bibinfo {volume} {71}},\
  \bibinfo {pages} {205107} (\bibinfo {year} {2005})}\BibitemShut {NoStop}%
\bibitem [{\citenamefont {Pittalis}\ \emph {et~al.}(2007)\citenamefont
  {Pittalis}, \citenamefont {Kurth}, \citenamefont {Sharma},\ and\
  \citenamefont {Gross}}]{Pittalis.Kurth.ea:Orbital.2007}%
  \BibitemOpen
  \bibfield  {author} {\bibinfo {author} {\bibfnamefont {S.}~\bibnamefont
  {Pittalis}}, \bibinfo {author} {\bibfnamefont {S.}~\bibnamefont {Kurth}},
  \bibinfo {author} {\bibfnamefont {S.}~\bibnamefont {Sharma}}, \ and\ \bibinfo
  {author} {\bibfnamefont {E.~K.~U.}\ \bibnamefont {Gross}},\ }\href {\doibase
  10.1063/1.2777140} {\bibfield  {journal} {\bibinfo  {journal} {J. Chem.
  Phys.}\ }\textbf {\bibinfo {volume} {127}},\ \bibinfo {pages} {124103}
  (\bibinfo {year} {2007})}\BibitemShut {NoStop}%
\bibitem [{\citenamefont {R\"as\"anen}, \citenamefont {Pittalis},\ and\
  \citenamefont {Proetto}(2010)}]{Rasanen.Pittalis.ea:Universal.2010}%
  \BibitemOpen
  \bibfield  {author} {\bibinfo {author} {\bibfnamefont {E.}~\bibnamefont
  {R\"as\"anen}}, \bibinfo {author} {\bibfnamefont {S.}~\bibnamefont
  {Pittalis}}, \ and\ \bibinfo {author} {\bibfnamefont {C.~R.}\ \bibnamefont
  {Proetto}},\ }\href {\doibase 10.1063/1.3300063} {\bibfield  {journal}
  {\bibinfo  {journal} {J. Chem. Phys.}\ }\textbf {\bibinfo {volume} {132}},\
  \bibinfo {pages} {044112} (\bibinfo {year} {2010})}\BibitemShut {NoStop}%
\bibitem [{\citenamefont {Pittalis}, \citenamefont {R\"as\"anen},\ and\
  \citenamefont {Gross}(2009)}]{Pittalis.Rasanen.ea:Gaussian.2009}%
  \BibitemOpen
  \bibfield  {author} {\bibinfo {author} {\bibfnamefont {S.}~\bibnamefont
  {Pittalis}}, \bibinfo {author} {\bibfnamefont {E.}~\bibnamefont
  {R\"as\"anen}}, \ and\ \bibinfo {author} {\bibfnamefont {E.~K.~U.}\
  \bibnamefont {Gross}},\ }\href {\doibase 10.1103/PhysRevA.80.032515}
  {\bibfield  {journal} {\bibinfo  {journal} {Phys. Rev. A}\ }\textbf {\bibinfo
  {volume} {80}},\ \bibinfo {pages} {032515} (\bibinfo {year}
  {2009})}\BibitemShut {NoStop}%
\bibitem [{\citenamefont {Bates}\ and\ \citenamefont
  {Furche}(2012)}]{Bates.Furche:Harnessing.2012}%
  \BibitemOpen
  \bibfield  {author} {\bibinfo {author} {\bibfnamefont {J.~E.}\ \bibnamefont
  {Bates}}\ and\ \bibinfo {author} {\bibfnamefont {F.}~\bibnamefont {Furche}},\
  }\href {\doibase 10.1063/1.4759080} {\bibfield  {journal} {\bibinfo
  {journal} {J. Chem. Phys.}\ }\textbf {\bibinfo {volume} {137}},\ \bibinfo
  {pages} {164105} (\bibinfo {year} {2012})}\BibitemShut {NoStop}%
\bibitem [{\citenamefont {Maier}, \citenamefont {Ikabata},\ and\ \citenamefont
  {Nakai}(2020)}]{Maier.Ikabata.ea:Relativistic.2020}%
  \BibitemOpen
  \bibfield  {author} {\bibinfo {author} {\bibfnamefont {T.~M.}\ \bibnamefont
  {Maier}}, \bibinfo {author} {\bibfnamefont {Y.}~\bibnamefont {Ikabata}}, \
  and\ \bibinfo {author} {\bibfnamefont {H.}~\bibnamefont {Nakai}},\ }\href
  {\doibase 10.1063/5.0010400} {\bibfield  {journal} {\bibinfo  {journal} {J.
  Chem. Phys.}\ }\textbf {\bibinfo {volume} {152}},\ \bibinfo {pages} {214103}
  (\bibinfo {year} {2020})}\BibitemShut {NoStop}%
\bibitem [{\citenamefont {Holzer}, \citenamefont {Franzke},\ and\ \citenamefont
  {Pausch}(2022)}]{Holzer.Franke.ea:Current.2022}%
  \BibitemOpen
  \bibfield  {author} {\bibinfo {author} {\bibfnamefont {C.}~\bibnamefont
  {Holzer}}, \bibinfo {author} {\bibfnamefont {Y.~J.}\ \bibnamefont {Franzke}},
  \ and\ \bibinfo {author} {\bibfnamefont {A.}~\bibnamefont {Pausch}},\ }\href
  {\doibase 10.1063/5.0122394} {\bibfield  {journal} {\bibinfo  {journal} {J.
  Chem. Phys.}\ }\textbf {\bibinfo {volume} {157}},\ \bibinfo {pages} {204102}
  (\bibinfo {year} {2022})}\BibitemShut {NoStop}%
\bibitem [{\citenamefont {Desmarais}\ \emph
  {et~al.}(2024{\natexlab{a}})\citenamefont {Desmarais}, \citenamefont
  {Ambrogio}, \citenamefont {Vignale}, \citenamefont {Erba},\ and\
  \citenamefont {Pittalis}}]{Desmarais.Ambrogio.ea:Generalized.2024}%
  \BibitemOpen
  \bibfield  {author} {\bibinfo {author} {\bibfnamefont {J.~K.}\ \bibnamefont
  {Desmarais}}, \bibinfo {author} {\bibfnamefont {G.}~\bibnamefont {Ambrogio}},
  \bibinfo {author} {\bibfnamefont {G.}~\bibnamefont {Vignale}}, \bibinfo
  {author} {\bibfnamefont {A.}~\bibnamefont {Erba}}, \ and\ \bibinfo {author}
  {\bibfnamefont {S.}~\bibnamefont {Pittalis}},\ }\href {\doibase
  10.1103/PhysRevMaterials.8.013802} {\bibfield  {journal} {\bibinfo  {journal}
  {Phys. Rev. Mater.}\ }\textbf {\bibinfo {volume} {8}},\ \bibinfo {pages}
  {013802} (\bibinfo {year} {2024}{\natexlab{a}})}\BibitemShut {NoStop}%
\bibitem [{\citenamefont {Desmarais}\ \emph
  {et~al.}(2024{\natexlab{b}})\citenamefont {Desmarais}, \citenamefont {Maul},
  \citenamefont {Civalleri}, \citenamefont {Erba}, \citenamefont {Vignale},\
  and\ \citenamefont {Pittalis}}]{Desmarais.Maul.ea:2024}%
  \BibitemOpen
  \bibfield  {author} {\bibinfo {author} {\bibfnamefont {J.~K.}\ \bibnamefont
  {Desmarais}}, \bibinfo {author} {\bibfnamefont {J.}~\bibnamefont {Maul}},
  \bibinfo {author} {\bibfnamefont {B.}~\bibnamefont {Civalleri}}, \bibinfo
  {author} {\bibfnamefont {A.}~\bibnamefont {Erba}}, \bibinfo {author}
  {\bibfnamefont {G.}~\bibnamefont {Vignale}}, \ and\ \bibinfo {author}
  {\bibfnamefont {S.}~\bibnamefont {Pittalis}},\ }\href {\doibase
  10.48550/arXiv.2401.07581} {\bibfield  {journal} {\bibinfo  {journal}
  {arXiv}\ } (\bibinfo {year} {2024}{\natexlab{b}}),\
  10.48550/arXiv.2401.07581}\BibitemShut {NoStop}%
\bibitem [{\citenamefont {Tellgren}\ \emph {et~al.}(2012)\citenamefont
  {Tellgren}, \citenamefont {Kvaal}, \citenamefont {Sagvolden}, \citenamefont
  {Ekstr\"om}, \citenamefont {Teale},\ and\ \citenamefont
  {Helgaker}}]{Tellgren.Kvaal.ea:Choice.2012}%
  \BibitemOpen
  \bibfield  {author} {\bibinfo {author} {\bibfnamefont {E.~I.}\ \bibnamefont
  {Tellgren}}, \bibinfo {author} {\bibfnamefont {S.}~\bibnamefont {Kvaal}},
  \bibinfo {author} {\bibfnamefont {E.}~\bibnamefont {Sagvolden}}, \bibinfo
  {author} {\bibfnamefont {U.}~\bibnamefont {Ekstr\"om}}, \bibinfo {author}
  {\bibfnamefont {A.~M.}\ \bibnamefont {Teale}}, \ and\ \bibinfo {author}
  {\bibfnamefont {T.}~\bibnamefont {Helgaker}},\ }\href {\doibase
  10.1103/PhysRevA.86.062506} {\bibfield  {journal} {\bibinfo  {journal} {Phys.
  Rev. A}\ }\textbf {\bibinfo {volume} {86}},\ \bibinfo {pages} {062506}
  (\bibinfo {year} {2012})}\BibitemShut {NoStop}%
\bibitem [{\citenamefont {Becke}(2002)}]{Becke:Current.2002}%
  \BibitemOpen
  \bibfield  {author} {\bibinfo {author} {\bibfnamefont {A.~D.}\ \bibnamefont
  {Becke}},\ }\href {\doibase 10.1063/1.1503772} {\bibfield  {journal}
  {\bibinfo  {journal} {J. Chem. Phys.}\ }\textbf {\bibinfo {volume} {117}},\
  \bibinfo {pages} {6935} (\bibinfo {year} {2002})}\BibitemShut {NoStop}%
\bibitem [{\citenamefont {K\"ubler}\ \emph {et~al.}(1988)\citenamefont
  {K\"ubler}, \citenamefont {H\"ock}, \citenamefont {Sticht},\ and\
  \citenamefont {Williams}}]{Kubler.Hock.ea:Density.1988}%
  \BibitemOpen
  \bibfield  {author} {\bibinfo {author} {\bibfnamefont {J.}~\bibnamefont
  {K\"ubler}}, \bibinfo {author} {\bibfnamefont {K.-H.}\ \bibnamefont
  {H\"ock}}, \bibinfo {author} {\bibfnamefont {J.}~\bibnamefont {Sticht}}, \
  and\ \bibinfo {author} {\bibfnamefont {A.~R.}\ \bibnamefont {Williams}},\
  }\href {\doibase 10.1088/0305-4608/18/3/018} {\bibfield  {journal} {\bibinfo
  {journal} {J. Phys. F Metal Phys.}\ }\textbf {\bibinfo {volume} {18}},\
  \bibinfo {pages} {469} (\bibinfo {year} {1988})}\BibitemShut {NoStop}%
\bibitem [{\citenamefont {Van~W{\"u}llen}(2002)}]{Van-Wullen:Spin.2002}%
  \BibitemOpen
  \bibfield  {author} {\bibinfo {author} {\bibfnamefont {C.}~\bibnamefont
  {Van~W{\"u}llen}},\ }\href {\doibase 10.1002/jcc.10043} {\bibfield  {journal}
  {\bibinfo  {journal} {J. Comput. Chem.}\ }\textbf {\bibinfo {volume} {23}},\
  \bibinfo {pages} {779} (\bibinfo {year} {2002})}\BibitemShut {NoStop}%
\bibitem [{\citenamefont {Saue}\ and\ \citenamefont
  {Helgaker}(2002)}]{Saue.Helgaker:Four-component.2002}%
  \BibitemOpen
  \bibfield  {author} {\bibinfo {author} {\bibfnamefont {T.}~\bibnamefont
  {Saue}}\ and\ \bibinfo {author} {\bibfnamefont {T.}~\bibnamefont
  {Helgaker}},\ }\href {\doibase 10.1002/jcc.10066} {\bibfield  {journal}
  {\bibinfo  {journal} {J. Comput. Chem.}\ }\textbf {\bibinfo {volume} {23}},\
  \bibinfo {pages} {814} (\bibinfo {year} {2002})}\BibitemShut {NoStop}%
\bibitem [{\citenamefont {Armbruster}\ \emph {et~al.}(2008)\citenamefont
  {Armbruster}, \citenamefont {Weigend}, \citenamefont {van W{\"u}llen},\ and\
  \citenamefont {Klopper}}]{Armbruster.Weigend.ea:Self-consistent.2008}%
  \BibitemOpen
  \bibfield  {author} {\bibinfo {author} {\bibfnamefont {M.~K.}\ \bibnamefont
  {Armbruster}}, \bibinfo {author} {\bibfnamefont {F.}~\bibnamefont {Weigend}},
  \bibinfo {author} {\bibfnamefont {C.}~\bibnamefont {van W{\"u}llen}}, \ and\
  \bibinfo {author} {\bibfnamefont {W.}~\bibnamefont {Klopper}},\ }\href
  {\doibase 10.1039/B717719D} {\bibfield  {journal} {\bibinfo  {journal} {Phys.
  Chem. Chem. Phys.}\ }\textbf {\bibinfo {volume} {10}},\ \bibinfo {pages}
  {1748} (\bibinfo {year} {2008})}\BibitemShut {NoStop}%
\bibitem [{\citenamefont {Peralta}, \citenamefont {Scuseria},\ and\
  \citenamefont {Frisch}(2007)}]{Peralta.Scuseria.ea:Noncollinear.2007}%
  \BibitemOpen
  \bibfield  {author} {\bibinfo {author} {\bibfnamefont {J.~E.}\ \bibnamefont
  {Peralta}}, \bibinfo {author} {\bibfnamefont {G.~E.}\ \bibnamefont
  {Scuseria}}, \ and\ \bibinfo {author} {\bibfnamefont {M.~J.}\ \bibnamefont
  {Frisch}},\ }\href {\doibase 10.1103/PhysRevB.75.125119} {\bibfield
  {journal} {\bibinfo  {journal} {Phys. Rev. B}\ }\textbf {\bibinfo {volume}
  {75}},\ \bibinfo {pages} {125119} (\bibinfo {year} {2007})}\BibitemShut
  {NoStop}%
\bibitem [{\citenamefont {Scalmani}\ and\ \citenamefont
  {Frisch}(2012)}]{Scalmani.Frisch:New.2012}%
  \BibitemOpen
  \bibfield  {author} {\bibinfo {author} {\bibfnamefont {G.}~\bibnamefont
  {Scalmani}}\ and\ \bibinfo {author} {\bibfnamefont {M.~J.}\ \bibnamefont
  {Frisch}},\ }\href {\doibase 10.1021/ct300441z} {\bibfield  {journal}
  {\bibinfo  {journal} {J. Chem. Theory Comput.}\ }\textbf {\bibinfo {volume}
  {8}},\ \bibinfo {pages} {2193} (\bibinfo {year} {2012})}\BibitemShut
  {NoStop}%
\bibitem [{\citenamefont {Bulik}\ \emph {et~al.}(2013)\citenamefont {Bulik},
  \citenamefont {Scalmani}, \citenamefont {Frisch},\ and\ \citenamefont
  {Scuseria}}]{Bulik.Scalmani.ea:Noncollinear.2013}%
  \BibitemOpen
  \bibfield  {author} {\bibinfo {author} {\bibfnamefont {I.~W.}\ \bibnamefont
  {Bulik}}, \bibinfo {author} {\bibfnamefont {G.}~\bibnamefont {Scalmani}},
  \bibinfo {author} {\bibfnamefont {M.~J.}\ \bibnamefont {Frisch}}, \ and\
  \bibinfo {author} {\bibfnamefont {G.~E.}\ \bibnamefont {Scuseria}},\ }\href
  {\doibase 10.1103/PhysRevB.87.035117} {\bibfield  {journal} {\bibinfo
  {journal} {Phys. Rev. B}\ }\textbf {\bibinfo {volume} {87}},\ \bibinfo
  {pages} {035117} (\bibinfo {year} {2013})}\BibitemShut {NoStop}%
\bibitem [{\citenamefont {Baldes}\ and\ \citenamefont
  {Weigend}(2013)}]{Baldes.Weigend:Efficient.2013}%
  \BibitemOpen
  \bibfield  {author} {\bibinfo {author} {\bibfnamefont {A.}~\bibnamefont
  {Baldes}}\ and\ \bibinfo {author} {\bibfnamefont {F.}~\bibnamefont
  {Weigend}},\ }\href {\doibase 10.1080/00268976.2013.802037} {\bibfield
  {journal} {\bibinfo  {journal} {Mol. Phys.}\ }\textbf {\bibinfo {volume}
  {111}},\ \bibinfo {pages} {2617} (\bibinfo {year} {2013})}\BibitemShut
  {NoStop}%
\bibitem [{\citenamefont {Egidi}\ \emph {et~al.}(2017)\citenamefont {Egidi},
  \citenamefont {Sun}, \citenamefont {Goings}, \citenamefont {Scalmani},
  \citenamefont {Frisch},\ and\ \citenamefont
  {Li}}]{Egidi.Sun.ea:Two-Component.2017}%
  \BibitemOpen
  \bibfield  {author} {\bibinfo {author} {\bibfnamefont {F.}~\bibnamefont
  {Egidi}}, \bibinfo {author} {\bibfnamefont {S.}~\bibnamefont {Sun}}, \bibinfo
  {author} {\bibfnamefont {J.~J.}\ \bibnamefont {Goings}}, \bibinfo {author}
  {\bibfnamefont {G.}~\bibnamefont {Scalmani}}, \bibinfo {author}
  {\bibfnamefont {M.~J.}\ \bibnamefont {Frisch}}, \ and\ \bibinfo {author}
  {\bibfnamefont {X.}~\bibnamefont {Li}},\ }\href {\doibase
  10.1021/acs.jctc.7b00104} {\bibfield  {journal} {\bibinfo  {journal} {J.
  Chem. Theory Comput.}\ }\textbf {\bibinfo {volume} {13}},\ \bibinfo {pages}
  {2591} (\bibinfo {year} {2017})}\BibitemShut {NoStop}%
\bibitem [{\citenamefont {Komorovsky}, \citenamefont {Cherry},\ and\
  \citenamefont {Repisky}(2019)}]{Komorovsky.Cherry.ea:Four-component.2019}%
  \BibitemOpen
  \bibfield  {author} {\bibinfo {author} {\bibfnamefont {S.}~\bibnamefont
  {Komorovsky}}, \bibinfo {author} {\bibfnamefont {P.~J.}\ \bibnamefont
  {Cherry}}, \ and\ \bibinfo {author} {\bibfnamefont {M.}~\bibnamefont
  {Repisky}},\ }\href {\doibase 10.1063/1.5121713} {\bibfield  {journal}
  {\bibinfo  {journal} {J. Chem. Phys.}\ }\textbf {\bibinfo {volume} {151}},\
  \bibinfo {pages} {184111} (\bibinfo {year} {2019})}\BibitemShut {NoStop}%
\bibitem [{\citenamefont {Desmarais}\ \emph {et~al.}(2021)\citenamefont
  {Desmarais}, \citenamefont {Komorovsky}, \citenamefont {Flament},\ and\
  \citenamefont {Erba}}]{Desmarais.Komorovsky.ea:Spin-orbit.2021}%
  \BibitemOpen
  \bibfield  {author} {\bibinfo {author} {\bibfnamefont {J.~K.}\ \bibnamefont
  {Desmarais}}, \bibinfo {author} {\bibfnamefont {S.}~\bibnamefont
  {Komorovsky}}, \bibinfo {author} {\bibfnamefont {J.-P.}\ \bibnamefont
  {Flament}}, \ and\ \bibinfo {author} {\bibfnamefont {A.}~\bibnamefont
  {Erba}},\ }\href {\doibase 10.1063/5.0051447} {\bibfield  {journal} {\bibinfo
   {journal} {J. Chem. Phys.}\ }\textbf {\bibinfo {volume} {154}},\ \bibinfo
  {pages} {204110} (\bibinfo {year} {2021})}\BibitemShut {NoStop}%
\bibitem [{\citenamefont {Bruder}, \citenamefont {Franzke},\ and\ \citenamefont
  {Weigend}(2022)}]{Bruder.Franzke.ea:Paramagnetic.2022}%
  \BibitemOpen
  \bibfield  {author} {\bibinfo {author} {\bibfnamefont {F.}~\bibnamefont
  {Bruder}}, \bibinfo {author} {\bibfnamefont {Y.~J.}\ \bibnamefont {Franzke}},
  \ and\ \bibinfo {author} {\bibfnamefont {F.}~\bibnamefont {Weigend}},\ }\href
  {\doibase 10.1021/acs.jpca.2c03579} {\bibfield  {journal} {\bibinfo
  {journal} {J. Phys. Chem. A}\ }\textbf {\bibinfo {volume} {126}},\ \bibinfo
  {pages} {5050} (\bibinfo {year} {2022})}\BibitemShut {NoStop}%
\bibitem [{\citenamefont {Bruder}\ \emph {et~al.}(2023)\citenamefont {Bruder},
  \citenamefont {Franzke}, \citenamefont {Holzer},\ and\ \citenamefont
  {Weigend}}]{Bruder.Franzke.ea:Zero-Field.2023}%
  \BibitemOpen
  \bibfield  {author} {\bibinfo {author} {\bibfnamefont {F.}~\bibnamefont
  {Bruder}}, \bibinfo {author} {\bibfnamefont {Y.~J.}\ \bibnamefont {Franzke}},
  \bibinfo {author} {\bibfnamefont {C.}~\bibnamefont {Holzer}}, \ and\ \bibinfo
  {author} {\bibfnamefont {F.}~\bibnamefont {Weigend}},\ }\href {\doibase
  10.1063/5.0175758} {\bibfield  {journal} {\bibinfo  {journal} {J. Chem.
  Phys.}\ }\textbf {\bibinfo {volume} {159}},\ \bibinfo {pages} {194117}
  (\bibinfo {year} {2023})}\BibitemShut {NoStop}%
\bibitem [{\citenamefont {Franzke}\ \emph {et~al.}(2024)\citenamefont
  {Franzke}, \citenamefont {Bruder}, \citenamefont {Gillhuber}, \citenamefont
  {Holzer},\ and\ \citenamefont
  {Weigend}}]{Franzke.Bruder.ea:Paramagnetic.2024}%
  \BibitemOpen
  \bibfield  {author} {\bibinfo {author} {\bibfnamefont {Y.~J.}\ \bibnamefont
  {Franzke}}, \bibinfo {author} {\bibfnamefont {F.}~\bibnamefont {Bruder}},
  \bibinfo {author} {\bibfnamefont {S.}~\bibnamefont {Gillhuber}}, \bibinfo
  {author} {\bibfnamefont {C.}~\bibnamefont {Holzer}}, \ and\ \bibinfo {author}
  {\bibfnamefont {F.}~\bibnamefont {Weigend}},\ }\href {\doibase
  10.1021/acs.jpca.3c07093} {\bibfield  {journal} {\bibinfo  {journal} {J.
  Phys. Chem. A}\ }\textbf {\bibinfo {volume} {128}},\ \bibinfo {pages} {670}
  (\bibinfo {year} {2024})}\BibitemShut {NoStop}%
\bibitem [{\citenamefont {{TURBOMOLE
  GmbH}}(2024{\natexlab{a}})}]{TURBOMOLE-manual}%
  \BibitemOpen
  \bibfield  {author} {\bibinfo {author} {\bibnamefont {{TURBOMOLE GmbH}}},\
  }\href@noop {} {} (\bibinfo {year} {2024}{\natexlab{a}}),\ \bibinfo {note}
  {manual of {TURBOMOLE V7.8.1}, a development of {University of Karlsruhe} and
  {Forschungszentrum Karlsruhe GmbH}, 1989-2007, {TURBOMOLE GmbH}, since 2007;
  available from
  \url{https://www.turbomole.org/turbomole/turbomole-documentation/} (retrieved
  March 4, 2024).}\BibitemShut {Stop}%
\bibitem [{\citenamefont {Franzke}, \citenamefont {Schosser},\ and\
  \citenamefont {Pauly}(2024)}]{Franzke.Schosser.ea:Efficient.2024}%
  \BibitemOpen
  \bibfield  {author} {\bibinfo {author} {\bibfnamefont {Y.~J.}\ \bibnamefont
  {Franzke}}, \bibinfo {author} {\bibfnamefont {W.~M.}\ \bibnamefont
  {Schosser}}, \ and\ \bibinfo {author} {\bibfnamefont {F.}~\bibnamefont
  {Pauly}},\ }\href {\doibase 10.48550/arXiv.2305.03817} {\bibfield  {journal}
  {\bibinfo  {journal} {Phys. Rev. B}\ } (\bibinfo {year} {2024}),\
  10.48550/arXiv.2305.03817},\ \bibinfo {note} {accepted}\BibitemShut {NoStop}%
\bibitem [{\citenamefont {Kasper}\ \emph {et~al.}(2020)\citenamefont {Kasper},
  \citenamefont {Jenkins}, \citenamefont {Sun},\ and\ \citenamefont
  {Li}}]{Kasper.Jenkins.ea:Perspective.2020}%
  \BibitemOpen
  \bibfield  {author} {\bibinfo {author} {\bibfnamefont {J.~M.}\ \bibnamefont
  {Kasper}}, \bibinfo {author} {\bibfnamefont {A.~J.}\ \bibnamefont {Jenkins}},
  \bibinfo {author} {\bibfnamefont {S.}~\bibnamefont {Sun}}, \ and\ \bibinfo
  {author} {\bibfnamefont {X.}~\bibnamefont {Li}},\ }\href {\doibase
  10.1063/5.0015279} {\bibfield  {journal} {\bibinfo  {journal} {J. Chem.
  Phys.}\ }\textbf {\bibinfo {volume} {153}},\ \bibinfo {pages} {090903}
  (\bibinfo {year} {2020})}\BibitemShut {NoStop}%
\bibitem [{\citenamefont {Burow}, \citenamefont {Sierka},\ and\ \citenamefont
  {Mohamed}(2009)}]{Burow.Sierka.ea:Resolution.2009}%
  \BibitemOpen
  \bibfield  {author} {\bibinfo {author} {\bibfnamefont {A.~M.}\ \bibnamefont
  {Burow}}, \bibinfo {author} {\bibfnamefont {M.}~\bibnamefont {Sierka}}, \
  and\ \bibinfo {author} {\bibfnamefont {F.}~\bibnamefont {Mohamed}},\ }\href
  {\doibase 10.1063/1.3267858} {\bibfield  {journal} {\bibinfo  {journal} {J.
  Chem. Phys.}\ }\textbf {\bibinfo {volume} {131}},\ \bibinfo {pages} {214101}
  (\bibinfo {year} {2009})}\BibitemShut {NoStop}%
\bibitem [{\citenamefont {Burow}\ and\ \citenamefont
  {Sierka}(2011)}]{Burow.Sierka:Linear.2011}%
  \BibitemOpen
  \bibfield  {author} {\bibinfo {author} {\bibfnamefont {A.~M.}\ \bibnamefont
  {Burow}}\ and\ \bibinfo {author} {\bibfnamefont {M.}~\bibnamefont {Sierka}},\
  }\href {\doibase 10.1021/ct200412r} {\bibfield  {journal} {\bibinfo
  {journal} {J. Chem. Theory Comput.}\ }\textbf {\bibinfo {volume} {7}},\
  \bibinfo {pages} {3097} (\bibinfo {year} {2011})}\BibitemShut {NoStop}%
\bibitem [{\citenamefont {\L{}azarski}, \citenamefont {Burow},\ and\
  \citenamefont {Sierka}(2015)}]{Lazarski.Burow.ea:Density.2015}%
  \BibitemOpen
  \bibfield  {author} {\bibinfo {author} {\bibfnamefont {R.}~\bibnamefont
  {\L{}azarski}}, \bibinfo {author} {\bibfnamefont {A.~M.}\ \bibnamefont
  {Burow}}, \ and\ \bibinfo {author} {\bibfnamefont {M.}~\bibnamefont
  {Sierka}},\ }\href {\doibase 10.1021/acs.jctc.5b00252} {\bibfield  {journal}
  {\bibinfo  {journal} {J. Chem. Theory Comput.}\ }\textbf {\bibinfo {volume}
  {11}},\ \bibinfo {pages} {3029} (\bibinfo {year} {2015})}\BibitemShut
  {NoStop}%
\bibitem [{\citenamefont {\L{}azarski}\ \emph {et~al.}(2016)\citenamefont
  {\L{}azarski}, \citenamefont {Burow}, \citenamefont {Grajciar},\ and\
  \citenamefont {Sierka}}]{Lazarski.Burow.ea:Density.2016}%
  \BibitemOpen
  \bibfield  {author} {\bibinfo {author} {\bibfnamefont {R.}~\bibnamefont
  {\L{}azarski}}, \bibinfo {author} {\bibfnamefont {A.~M.}\ \bibnamefont
  {Burow}}, \bibinfo {author} {\bibfnamefont {L.}~\bibnamefont {Grajciar}}, \
  and\ \bibinfo {author} {\bibfnamefont {M.}~\bibnamefont {Sierka}},\ }\href
  {\doibase 10.1002/jcc.24477} {\bibfield  {journal} {\bibinfo  {journal} {J.
  Comput. Chem.}\ }\textbf {\bibinfo {volume} {37}},\ \bibinfo {pages} {2518}
  (\bibinfo {year} {2016})}\BibitemShut {NoStop}%
\bibitem [{\citenamefont {Grajciar}(2015)}]{Grajciar:Low.2015}%
  \BibitemOpen
  \bibfield  {author} {\bibinfo {author} {\bibfnamefont {L.}~\bibnamefont
  {Grajciar}},\ }\href {\doibase 10.1002/jcc.23961} {\bibfield  {journal}
  {\bibinfo  {journal} {J. Comput. Chem.}\ }\textbf {\bibinfo {volume} {36}},\
  \bibinfo {pages} {1521} (\bibinfo {year} {2015})}\BibitemShut {NoStop}%
\bibitem [{\citenamefont {Becker}\ and\ \citenamefont
  {Sierka}(2019)}]{Becker.Sierka:Density.2019}%
  \BibitemOpen
  \bibfield  {author} {\bibinfo {author} {\bibfnamefont {M.}~\bibnamefont
  {Becker}}\ and\ \bibinfo {author} {\bibfnamefont {M.}~\bibnamefont
  {Sierka}},\ }\href {\doibase 10.1002/jcc.26033} {\bibfield  {journal}
  {\bibinfo  {journal} {J. Comput. Chem.}\ }\textbf {\bibinfo {volume} {40}},\
  \bibinfo {pages} {2563} (\bibinfo {year} {2019})}\BibitemShut {NoStop}%
\bibitem [{\citenamefont {Irmler}, \citenamefont {Burow},\ and\ \citenamefont
  {Pauly}(2018)}]{Irmler.Burow.ea:Robust.2018}%
  \BibitemOpen
  \bibfield  {author} {\bibinfo {author} {\bibfnamefont {A.}~\bibnamefont
  {Irmler}}, \bibinfo {author} {\bibfnamefont {A.~M.}\ \bibnamefont {Burow}}, \
  and\ \bibinfo {author} {\bibfnamefont {F.}~\bibnamefont {Pauly}},\ }\href
  {\doibase 10.1021/acs.jctc.8b00122} {\bibfield  {journal} {\bibinfo
  {journal} {J. Chem. Theory Comput.}\ }\textbf {\bibinfo {volume} {14}},\
  \bibinfo {pages} {4567} (\bibinfo {year} {2018})}\BibitemShut {NoStop}%
\bibitem [{\citenamefont {Ahlrichs}\ \emph {et~al.}(1989)\citenamefont
  {Ahlrichs}, \citenamefont {B{\"a}r}, \citenamefont {H{\"a}ser}, \citenamefont
  {Horn},\ and\ \citenamefont {K{\"o}lmel}}]{Ahlrichs.Bar.ea:Electronic.1989}%
  \BibitemOpen
  \bibfield  {author} {\bibinfo {author} {\bibfnamefont {R.}~\bibnamefont
  {Ahlrichs}}, \bibinfo {author} {\bibfnamefont {M.}~\bibnamefont {B{\"a}r}},
  \bibinfo {author} {\bibfnamefont {M.}~\bibnamefont {H{\"a}ser}}, \bibinfo
  {author} {\bibfnamefont {H.}~\bibnamefont {Horn}}, \ and\ \bibinfo {author}
  {\bibfnamefont {C.}~\bibnamefont {K{\"o}lmel}},\ }\href {\doibase
  10.1016/0009-2614(89)85118-8} {\bibfield  {journal} {\bibinfo  {journal}
  {Chem. Phys. Lett.}\ }\textbf {\bibinfo {volume} {162}},\ \bibinfo {pages}
  {165} (\bibinfo {year} {1989})}\BibitemShut {NoStop}%
\bibitem [{\citenamefont {Franzke}\ \emph {et~al.}(2023)\citenamefont
  {Franzke}, \citenamefont {Holzer}, \citenamefont {Andersen}, \citenamefont
  {Begu\v{s}i\'c}, \citenamefont {Bruder}, \citenamefont {Coriani},
  \citenamefont {Della~Sala}, \citenamefont {Fabiano}, \citenamefont {Fedotov},
  \citenamefont {F\"urst}, \citenamefont {Gillhuber}, \citenamefont {Grotjahn},
  \citenamefont {Kaupp}, \citenamefont {Kehry}, \citenamefont {Krsti\'c},
  \citenamefont {Mack}, \citenamefont {Majumdar}, \citenamefont {Nguyen},
  \citenamefont {Parker}, \citenamefont {Pauly}, \citenamefont {Pausch},
  \citenamefont {Perlt}, \citenamefont {Phun}, \citenamefont {Rajabi},
  \citenamefont {Rappoport}, \citenamefont {Samal}, \citenamefont {Schrader},
  \citenamefont {Sharma}, \citenamefont {Tapavicza}, \citenamefont {Treß},
  \citenamefont {Voora}, \citenamefont {Wody{\'n}ski}, \citenamefont {Yu},
  \citenamefont {Zerulla}, \citenamefont {Furche}, \citenamefont {H\"attig},
  \citenamefont {Sierka}, \citenamefont {Tew},\ and\ \citenamefont
  {Weigend}}]{Franzke.Holzer.ea:TURBOMOLE.2023}%
  \BibitemOpen
  \bibfield  {author} {\bibinfo {author} {\bibfnamefont {Y.~J.}\ \bibnamefont
  {Franzke}}, \bibinfo {author} {\bibfnamefont {C.}~\bibnamefont {Holzer}},
  \bibinfo {author} {\bibfnamefont {J.~H.}\ \bibnamefont {Andersen}}, \bibinfo
  {author} {\bibfnamefont {T.}~\bibnamefont {Begu\v{s}i\'c}}, \bibinfo {author}
  {\bibfnamefont {F.}~\bibnamefont {Bruder}}, \bibinfo {author} {\bibfnamefont
  {S.}~\bibnamefont {Coriani}}, \bibinfo {author} {\bibfnamefont
  {F.}~\bibnamefont {Della~Sala}}, \bibinfo {author} {\bibfnamefont
  {E.}~\bibnamefont {Fabiano}}, \bibinfo {author} {\bibfnamefont {D.~A.}\
  \bibnamefont {Fedotov}}, \bibinfo {author} {\bibfnamefont {S.}~\bibnamefont
  {F\"urst}}, \bibinfo {author} {\bibfnamefont {S.}~\bibnamefont {Gillhuber}},
  \bibinfo {author} {\bibfnamefont {R.}~\bibnamefont {Grotjahn}}, \bibinfo
  {author} {\bibfnamefont {M.}~\bibnamefont {Kaupp}}, \bibinfo {author}
  {\bibfnamefont {M.}~\bibnamefont {Kehry}}, \bibinfo {author} {\bibfnamefont
  {M.}~\bibnamefont {Krsti\'c}}, \bibinfo {author} {\bibfnamefont
  {F.}~\bibnamefont {Mack}}, \bibinfo {author} {\bibfnamefont {S.}~\bibnamefont
  {Majumdar}}, \bibinfo {author} {\bibfnamefont {B.~D.}\ \bibnamefont
  {Nguyen}}, \bibinfo {author} {\bibfnamefont {S.~M.}\ \bibnamefont {Parker}},
  \bibinfo {author} {\bibfnamefont {F.}~\bibnamefont {Pauly}}, \bibinfo
  {author} {\bibfnamefont {A.}~\bibnamefont {Pausch}}, \bibinfo {author}
  {\bibfnamefont {E.}~\bibnamefont {Perlt}}, \bibinfo {author} {\bibfnamefont
  {G.~S.}\ \bibnamefont {Phun}}, \bibinfo {author} {\bibfnamefont
  {A.}~\bibnamefont {Rajabi}}, \bibinfo {author} {\bibfnamefont
  {D.}~\bibnamefont {Rappoport}}, \bibinfo {author} {\bibfnamefont
  {B.}~\bibnamefont {Samal}}, \bibinfo {author} {\bibfnamefont
  {T.}~\bibnamefont {Schrader}}, \bibinfo {author} {\bibfnamefont
  {M.}~\bibnamefont {Sharma}}, \bibinfo {author} {\bibfnamefont
  {E.}~\bibnamefont {Tapavicza}}, \bibinfo {author} {\bibfnamefont {R.~S.}\
  \bibnamefont {Treß}}, \bibinfo {author} {\bibfnamefont {V.}~\bibnamefont
  {Voora}}, \bibinfo {author} {\bibfnamefont {A.}~\bibnamefont {Wody{\'n}ski}},
  \bibinfo {author} {\bibfnamefont {J.~M.}\ \bibnamefont {Yu}}, \bibinfo
  {author} {\bibfnamefont {B.}~\bibnamefont {Zerulla}}, \bibinfo {author}
  {\bibfnamefont {F.}~\bibnamefont {Furche}}, \bibinfo {author} {\bibfnamefont
  {C.}~\bibnamefont {H\"attig}}, \bibinfo {author} {\bibfnamefont
  {M.}~\bibnamefont {Sierka}}, \bibinfo {author} {\bibfnamefont {D.~P.}\
  \bibnamefont {Tew}}, \ and\ \bibinfo {author} {\bibfnamefont
  {F.}~\bibnamefont {Weigend}},\ }\href {\doibase 10.1021/acs.jctc.3c00347}
  {\bibfield  {journal} {\bibinfo  {journal} {J. Chem. Theory Comput.}\
  }\textbf {\bibinfo {volume} {19}},\ \bibinfo {pages} {6859} (\bibinfo {year}
  {2023})}\BibitemShut {NoStop}%
\bibitem [{\citenamefont {{TURBOMOLE GmbH}}(2024{\natexlab{b}})}]{TURBOMOLE}%
  \BibitemOpen
  \bibfield  {author} {\bibinfo {author} {\bibnamefont {{TURBOMOLE GmbH}}},\
  }\href@noop {} {} (\bibinfo {year} {2024}{\natexlab{b}}),\ \bibinfo {note}
  {developers' version of {TURBOMOLE V7.8.1}, a development of {University of
  Karlsruhe} and {Forschungszentrum Karlsruhe GmbH}, 1989-2007, {TURBOMOLE
  GmbH}, since 2007; available from \url{https://www.turbomole.org} (retrieved
  March 4, 2024).}\BibitemShut {Stop}%
\bibitem [{\citenamefont {Stratmann}, \citenamefont {Scuseria},\ and\
  \citenamefont {Frisch}(1996)}]{Stratmann.Scuseria.ea:Achieving.1996}%
  \BibitemOpen
  \bibfield  {author} {\bibinfo {author} {\bibfnamefont {R.}~\bibnamefont
  {Stratmann}}, \bibinfo {author} {\bibfnamefont {G.~E.}\ \bibnamefont
  {Scuseria}}, \ and\ \bibinfo {author} {\bibfnamefont {M.~J.}\ \bibnamefont
  {Frisch}},\ }\href {\doibase https://doi.org/10.1016/0009-2614(96)00600-8}
  {\bibfield  {journal} {\bibinfo  {journal} {Chem. Phys.}\ }\textbf {\bibinfo
  {volume} {257}},\ \bibinfo {pages} {213} (\bibinfo {year}
  {1996})}\BibitemShut {NoStop}%
\bibitem [{\citenamefont {Delin}\ and\ \citenamefont
  {Tosatti}(2003)}]{Delin.Tosatti:Magnetic.2003}%
  \BibitemOpen
  \bibfield  {author} {\bibinfo {author} {\bibfnamefont {A.}~\bibnamefont
  {Delin}}\ and\ \bibinfo {author} {\bibfnamefont {E.}~\bibnamefont
  {Tosatti}},\ }\href {\doibase 10.1103/PhysRevB.68.144434} {\bibfield
  {journal} {\bibinfo  {journal} {Phys. Rev. B}\ }\textbf {\bibinfo {volume}
  {68}},\ \bibinfo {pages} {144434} (\bibinfo {year} {2003})}\BibitemShut
  {NoStop}%
\bibitem [{\citenamefont {Delin}\ and\ \citenamefont
  {Tosatti}(2004)}]{Delin.Tosatti:Emerging.2004}%
  \BibitemOpen
  \bibfield  {author} {\bibinfo {author} {\bibfnamefont {A.}~\bibnamefont
  {Delin}}\ and\ \bibinfo {author} {\bibfnamefont {E.}~\bibnamefont
  {Tosatti}},\ }\href {\doibase https://doi.org/10.1016/j.susc.2004.05.055}
  {\bibfield  {journal} {\bibinfo  {journal} {Surf. Sci.}\ }\textbf {\bibinfo
  {volume} {566--568}},\ \bibinfo {pages} {262} (\bibinfo {year}
  {2004})}\BibitemShut {NoStop}%
\bibitem [{\citenamefont {Fern\'andez-Rossier}\ \emph
  {et~al.}(2005)\citenamefont {Fern\'andez-Rossier}, \citenamefont {Jacob},
  \citenamefont {Untiedt},\ and\ \citenamefont
  {Palacios}}]{Fernandez-Rossier.Jacob.ea:Transport.2005}%
  \BibitemOpen
  \bibfield  {author} {\bibinfo {author} {\bibfnamefont {J.}~\bibnamefont
  {Fern\'andez-Rossier}}, \bibinfo {author} {\bibfnamefont {D.}~\bibnamefont
  {Jacob}}, \bibinfo {author} {\bibfnamefont {C.}~\bibnamefont {Untiedt}}, \
  and\ \bibinfo {author} {\bibfnamefont {J.~J.}\ \bibnamefont {Palacios}},\
  }\href {\doibase 10.1103/PhysRevB.72.224418} {\bibfield  {journal} {\bibinfo
  {journal} {Phys. Rev. B}\ }\textbf {\bibinfo {volume} {72}},\ \bibinfo
  {pages} {224418} (\bibinfo {year} {2005})}\BibitemShut {NoStop}%
\bibitem [{\citenamefont {Smogunov}\ \emph {et~al.}(2008)\citenamefont
  {Smogunov}, \citenamefont {Dal~Corso}, \citenamefont {Delin}, \citenamefont
  {Weht},\ and\ \citenamefont {Tosatti}}]{Smogunov.Dal-Corso.ea:Colossal.2008}%
  \BibitemOpen
  \bibfield  {author} {\bibinfo {author} {\bibfnamefont {A.}~\bibnamefont
  {Smogunov}}, \bibinfo {author} {\bibfnamefont {A.}~\bibnamefont {Dal~Corso}},
  \bibinfo {author} {\bibfnamefont {A.}~\bibnamefont {Delin}}, \bibinfo
  {author} {\bibfnamefont {R.}~\bibnamefont {Weht}}, \ and\ \bibinfo {author}
  {\bibfnamefont {E.}~\bibnamefont {Tosatti}},\ }\href {\doibase
  10.1038/nnano.2007.419} {\bibfield  {journal} {\bibinfo  {journal} {Nat.
  Nanotechnol.}\ }\textbf {\bibinfo {volume} {3}},\ \bibinfo {pages} {22}
  (\bibinfo {year} {2008})}\BibitemShut {NoStop}%
\bibitem [{\citenamefont {Garc{\'\i}a-Su{\'a}rez}\ \emph
  {et~al.}(2009)\citenamefont {Garc{\'\i}a-Su{\'a}rez}, \citenamefont
  {Manrique}, \citenamefont {Lambert},\ and\ \citenamefont
  {Ferrer}}]{Garca-Suarez.Manrique.ea:Anisotropic.2009}%
  \BibitemOpen
  \bibfield  {author} {\bibinfo {author} {\bibfnamefont {V.~M.}\ \bibnamefont
  {Garc{\'\i}a-Su{\'a}rez}}, \bibinfo {author} {\bibfnamefont {D.~Z.}\
  \bibnamefont {Manrique}}, \bibinfo {author} {\bibfnamefont {C.~J.}\
  \bibnamefont {Lambert}}, \ and\ \bibinfo {author} {\bibfnamefont
  {J.}~\bibnamefont {Ferrer}},\ }\href {\doibase 10.1103/PhysRevB.79.060408}
  {\bibfield  {journal} {\bibinfo  {journal} {Phys. Rev. B}\ }\textbf {\bibinfo
  {volume} {79}},\ \bibinfo {pages} {060408(R)} (\bibinfo {year}
  {2009})}\BibitemShut {NoStop}%
\bibitem [{\citenamefont {Tao}\ \emph {et~al.}(2003)\citenamefont {Tao},
  \citenamefont {Perdew}, \citenamefont {Staroverov},\ and\ \citenamefont
  {Scuseria}}]{Tao.Perdew.ea:Climbing.2003}%
  \BibitemOpen
  \bibfield  {author} {\bibinfo {author} {\bibfnamefont {J.}~\bibnamefont
  {Tao}}, \bibinfo {author} {\bibfnamefont {J.~P.}\ \bibnamefont {Perdew}},
  \bibinfo {author} {\bibfnamefont {V.~N.}\ \bibnamefont {Staroverov}}, \ and\
  \bibinfo {author} {\bibfnamefont {G.~E.}\ \bibnamefont {Scuseria}},\ }\href
  {\doibase 10.1103/PhysRevLett.91.146401} {\bibfield  {journal} {\bibinfo
  {journal} {Phys. Rev. Lett.}\ }\textbf {\bibinfo {volume} {91}},\ \bibinfo
  {pages} {146401} (\bibinfo {year} {2003})}\BibitemShut {NoStop}%
\bibitem [{\citenamefont {Weigend}\ and\ \citenamefont
  {Baldes}(2010)}]{Weigend.Baldes:Segmented.2010}%
  \BibitemOpen
  \bibfield  {author} {\bibinfo {author} {\bibfnamefont {F.}~\bibnamefont
  {Weigend}}\ and\ \bibinfo {author} {\bibfnamefont {A.}~\bibnamefont
  {Baldes}},\ }\href {\doibase 10.1063/1.3495681} {\bibfield  {journal}
  {\bibinfo  {journal} {J. Chem. Phys.}\ }\textbf {\bibinfo {volume} {133}},\
  \bibinfo {pages} {174102} (\bibinfo {year} {2010})}\BibitemShut {NoStop}%
\bibitem [{\citenamefont {Figgen}\ \emph {et~al.}(2009)\citenamefont {Figgen},
  \citenamefont {Peterson}, \citenamefont {Dolg},\ and\ \citenamefont
  {Stoll}}]{Figgen.Peterson.ea:Energy-consistent.2009}%
  \BibitemOpen
  \bibfield  {author} {\bibinfo {author} {\bibfnamefont {D.}~\bibnamefont
  {Figgen}}, \bibinfo {author} {\bibfnamefont {K.~A.}\ \bibnamefont
  {Peterson}}, \bibinfo {author} {\bibfnamefont {M.}~\bibnamefont {Dolg}}, \
  and\ \bibinfo {author} {\bibfnamefont {H.}~\bibnamefont {Stoll}},\ }\href
  {\doibase 10.1063/1.3119665} {\bibfield  {journal} {\bibinfo  {journal} {J.
  Chem. Phys.}\ }\textbf {\bibinfo {volume} {130}},\ \bibinfo {pages} {164108}
  (\bibinfo {year} {2009})}\BibitemShut {NoStop}%
\bibitem [{\citenamefont {Kresse}\ and\ \citenamefont
  {Furthm\"uller}(1996)}]{Kresse.Furthmuller:Efficiency.1996}%
  \BibitemOpen
  \bibfield  {author} {\bibinfo {author} {\bibfnamefont {G.}~\bibnamefont
  {Kresse}}\ and\ \bibinfo {author} {\bibfnamefont {J.}~\bibnamefont
  {Furthm\"uller}},\ }\href {\doibase
  https://doi.org/10.1016/0927-0256(96)00008-0} {\bibfield  {journal} {\bibinfo
   {journal} {Comput. Mater. Sci.}\ }\textbf {\bibinfo {volume} {6}},\ \bibinfo
  {pages} {15} (\bibinfo {year} {1996})}\BibitemShut {NoStop}%
\bibitem [{\citenamefont {Peintinger}, \citenamefont {Oliveira},\ and\
  \citenamefont {Bredow}(2013)}]{Peintinger.Oliveira.ea:Consistent.2013}%
  \BibitemOpen
  \bibfield  {author} {\bibinfo {author} {\bibfnamefont {M.~F.}\ \bibnamefont
  {Peintinger}}, \bibinfo {author} {\bibfnamefont {D.~V.}\ \bibnamefont
  {Oliveira}}, \ and\ \bibinfo {author} {\bibfnamefont {T.}~\bibnamefont
  {Bredow}},\ }\href {\doibase https://doi.org/10.1002/jcc.23153} {\bibfield
  {journal} {\bibinfo  {journal} {J. Comput. Chem.}\ }\textbf {\bibinfo
  {volume} {34}},\ \bibinfo {pages} {451} (\bibinfo {year} {2013})}\BibitemShut
  {NoStop}%
\bibitem [{\citenamefont {Laun}, \citenamefont {Vilela~Oliveira},\ and\
  \citenamefont {Bredow}(2018)}]{Laun.Vilela-Oliveira.ea:Consistent.2018}%
  \BibitemOpen
  \bibfield  {author} {\bibinfo {author} {\bibfnamefont {J.}~\bibnamefont
  {Laun}}, \bibinfo {author} {\bibfnamefont {D.}~\bibnamefont
  {Vilela~Oliveira}}, \ and\ \bibinfo {author} {\bibfnamefont {T.}~\bibnamefont
  {Bredow}},\ }\href {\doibase https://doi.org/10.1002/jcc.25195} {\bibfield
  {journal} {\bibinfo  {journal} {J. Comput. Chem.}\ }\textbf {\bibinfo
  {volume} {39}},\ \bibinfo {pages} {1285} (\bibinfo {year}
  {2018})}\BibitemShut {NoStop}%
\bibitem [{\citenamefont {Vilela~Oliveira}\ \emph {et~al.}(2019)\citenamefont
  {Vilela~Oliveira}, \citenamefont {Laun}, \citenamefont {Peintinger},\ and\
  \citenamefont {Bredow}}]{Vilela-Oliveira.Laun.ea:BSSE-correction.2019}%
  \BibitemOpen
  \bibfield  {author} {\bibinfo {author} {\bibfnamefont {D.}~\bibnamefont
  {Vilela~Oliveira}}, \bibinfo {author} {\bibfnamefont {J.}~\bibnamefont
  {Laun}}, \bibinfo {author} {\bibfnamefont {M.~F.}\ \bibnamefont
  {Peintinger}}, \ and\ \bibinfo {author} {\bibfnamefont {T.}~\bibnamefont
  {Bredow}},\ }\href {\doibase https://doi.org/10.1002/jcc.26013} {\bibfield
  {journal} {\bibinfo  {journal} {J. Comput. Chem.}\ }\textbf {\bibinfo
  {volume} {40}},\ \bibinfo {pages} {2364} (\bibinfo {year}
  {2019})}\BibitemShut {NoStop}%
\bibitem [{\citenamefont {Laun}\ and\ \citenamefont
  {Bredow}(2021)}]{Laun.Bredow:BSSE-corrected.2021}%
  \BibitemOpen
  \bibfield  {author} {\bibinfo {author} {\bibfnamefont {J.}~\bibnamefont
  {Laun}}\ and\ \bibinfo {author} {\bibfnamefont {T.}~\bibnamefont {Bredow}},\
  }\href {\doibase https://doi.org/10.1002/jcc.26521} {\bibfield  {journal}
  {\bibinfo  {journal} {J. Comput. Chem.}\ }\textbf {\bibinfo {volume} {42}},\
  \bibinfo {pages} {1064} (\bibinfo {year} {2021})}\BibitemShut {NoStop}%
\bibitem [{\citenamefont {Laun}\ and\ \citenamefont
  {Bredow}(2022)}]{Laun.Bredow:BSSE-corrected.2022}%
  \BibitemOpen
  \bibfield  {author} {\bibinfo {author} {\bibfnamefont {J.}~\bibnamefont
  {Laun}}\ and\ \bibinfo {author} {\bibfnamefont {T.}~\bibnamefont {Bredow}},\
  }\href {\doibase https://doi.org/10.1002/jcc.26839} {\bibfield  {journal}
  {\bibinfo  {journal} {J. Comput. Chem.}\ }\textbf {\bibinfo {volume} {43}},\
  \bibinfo {pages} {839} (\bibinfo {year} {2022})}\BibitemShut {NoStop}%
\bibitem [{\citenamefont {Seidler}, \citenamefont {Laun},\ and\ \citenamefont
  {Bredow}(2023)}]{Seidler.Laun.ea:BSSE-corrected.2023}%
  \BibitemOpen
  \bibfield  {author} {\bibinfo {author} {\bibfnamefont {L.~M.}\ \bibnamefont
  {Seidler}}, \bibinfo {author} {\bibfnamefont {J.}~\bibnamefont {Laun}}, \
  and\ \bibinfo {author} {\bibfnamefont {T.}~\bibnamefont {Bredow}},\ }\href
  {\doibase https://doi.org/10.1002/jcc.27097} {\bibfield  {journal} {\bibinfo
  {journal} {J. Comput. Chem.}\ }\textbf {\bibinfo {volume} {44}},\ \bibinfo
  {pages} {1418} (\bibinfo {year} {2023})}\BibitemShut {NoStop}%
\bibitem [{\citenamefont {Armbruster}, \citenamefont {Klopper},\ and\
  \citenamefont {Weigend}(2006)}]{Armbruster.Klopper.ea:Basis-set.2006}%
  \BibitemOpen
  \bibfield  {author} {\bibinfo {author} {\bibfnamefont {M.~K.}\ \bibnamefont
  {Armbruster}}, \bibinfo {author} {\bibfnamefont {W.}~\bibnamefont {Klopper}},
  \ and\ \bibinfo {author} {\bibfnamefont {F.}~\bibnamefont {Weigend}},\ }\href
  {\doibase 10.1039/B610211E} {\bibfield  {journal} {\bibinfo  {journal} {Phys.
  Chem. Chem. Phys.}\ }\textbf {\bibinfo {volume} {8}},\ \bibinfo {pages}
  {4862} (\bibinfo {year} {2006})}\BibitemShut {NoStop}%
\bibitem [{\citenamefont {Zhao}\ and\ \citenamefont
  {Truhlar}(2006)}]{Zhao.Truhlar:new.2006}%
  \BibitemOpen
  \bibfield  {author} {\bibinfo {author} {\bibfnamefont {Y.}~\bibnamefont
  {Zhao}}\ and\ \bibinfo {author} {\bibfnamefont {D.~G.}\ \bibnamefont
  {Truhlar}},\ }\href {\doibase 10.1063/1.2370993} {\bibfield  {journal}
  {\bibinfo  {journal} {J. Chem. Phys.}\ }\textbf {\bibinfo {volume} {125}},\
  \bibinfo {pages} {194101} (\bibinfo {year} {2006})}\BibitemShut {NoStop}%
\bibitem [{\citenamefont {Furness}\ \emph
  {et~al.}(2020{\natexlab{a}})\citenamefont {Furness}, \citenamefont {Kaplan},
  \citenamefont {Ning}, \citenamefont {Perdew},\ and\ \citenamefont
  {Sun}}]{Furness.Kaplan.ea:Accurate.2020}%
  \BibitemOpen
  \bibfield  {author} {\bibinfo {author} {\bibfnamefont {J.~W.}\ \bibnamefont
  {Furness}}, \bibinfo {author} {\bibfnamefont {A.~D.}\ \bibnamefont {Kaplan}},
  \bibinfo {author} {\bibfnamefont {J.}~\bibnamefont {Ning}}, \bibinfo {author}
  {\bibfnamefont {J.~P.}\ \bibnamefont {Perdew}}, \ and\ \bibinfo {author}
  {\bibfnamefont {J.}~\bibnamefont {Sun}},\ }\href {\doibase
  10.1021/acs.jpclett.0c02405} {\bibfield  {journal} {\bibinfo  {journal} {J.
  Phys. Chem. Lett.}\ }\textbf {\bibinfo {volume} {11}},\ \bibinfo {pages}
  {8208} (\bibinfo {year} {2020}{\natexlab{a}})}\BibitemShut {NoStop}%
\bibitem [{\citenamefont {Furness}\ \emph
  {et~al.}(2020{\natexlab{b}})\citenamefont {Furness}, \citenamefont {Kaplan},
  \citenamefont {Ning}, \citenamefont {Perdew},\ and\ \citenamefont
  {Sun}}]{Furness.Kaplan.ea:Correction.2020}%
  \BibitemOpen
  \bibfield  {author} {\bibinfo {author} {\bibfnamefont {J.~W.}\ \bibnamefont
  {Furness}}, \bibinfo {author} {\bibfnamefont {A.~D.}\ \bibnamefont {Kaplan}},
  \bibinfo {author} {\bibfnamefont {J.}~\bibnamefont {Ning}}, \bibinfo {author}
  {\bibfnamefont {J.~P.}\ \bibnamefont {Perdew}}, \ and\ \bibinfo {author}
  {\bibfnamefont {J.}~\bibnamefont {Sun}},\ }\href {\doibase
  10.1021/acs.jpclett.0c03077} {\bibfield  {journal} {\bibinfo  {journal} {J.
  Phys. Chem. Lett.}\ }\textbf {\bibinfo {volume} {11}},\ \bibinfo {pages}
  {9248} (\bibinfo {year} {2020}{\natexlab{b}})}\BibitemShut {NoStop}%
\bibitem [{\citenamefont {Aschebrock}\ and\ \citenamefont
  {K\"ummel}(2019)}]{Aschebrock.Kummel:Ultranonlocality.2019}%
  \BibitemOpen
  \bibfield  {author} {\bibinfo {author} {\bibfnamefont {T.}~\bibnamefont
  {Aschebrock}}\ and\ \bibinfo {author} {\bibfnamefont {S.}~\bibnamefont
  {K\"ummel}},\ }\href {\doibase 10.1103/PhysRevResearch.1.033082} {\bibfield
  {journal} {\bibinfo  {journal} {Phys. Rev. Res.}\ }\textbf {\bibinfo {volume}
  {1}},\ \bibinfo {pages} {033082} (\bibinfo {year} {2019})}\BibitemShut
  {NoStop}%
\bibitem [{\citenamefont {Perdew}\ \emph {et~al.}(1999)\citenamefont {Perdew},
  \citenamefont {Kurth}, \citenamefont {Zupan},\ and\ \citenamefont
  {Blaha}}]{Perdew.Kurth.ea:Accurate.1999}%
  \BibitemOpen
  \bibfield  {author} {\bibinfo {author} {\bibfnamefont {J.~P.}\ \bibnamefont
  {Perdew}}, \bibinfo {author} {\bibfnamefont {S.}~\bibnamefont {Kurth}},
  \bibinfo {author} {\bibfnamefont {A.~c.~v.}\ \bibnamefont {Zupan}}, \ and\
  \bibinfo {author} {\bibfnamefont {P.}~\bibnamefont {Blaha}},\ }\href
  {\doibase 10.1103/PhysRevLett.82.2544} {\bibfield  {journal} {\bibinfo
  {journal} {Phys. Rev. Lett.}\ }\textbf {\bibinfo {volume} {82}},\ \bibinfo
  {pages} {2544} (\bibinfo {year} {1999})}\BibitemShut {NoStop}%
\bibitem [{\citenamefont {Perdew}, \citenamefont {Burke},\ and\ \citenamefont
  {Ernzerhof}(1996)}]{Perdew.Burke.ea:Generalized.1996}%
  \BibitemOpen
  \bibfield  {author} {\bibinfo {author} {\bibfnamefont {J.~P.}\ \bibnamefont
  {Perdew}}, \bibinfo {author} {\bibfnamefont {K.}~\bibnamefont {Burke}}, \
  and\ \bibinfo {author} {\bibfnamefont {M.}~\bibnamefont {Ernzerhof}},\ }\href
  {\doibase 10.1103/PhysRevLett.77.3865} {\bibfield  {journal} {\bibinfo
  {journal} {Phys. Rev. Lett.}\ }\textbf {\bibinfo {volume} {77}},\ \bibinfo
  {pages} {3865} (\bibinfo {year} {1996})}\BibitemShut {NoStop}%
\bibitem [{\citenamefont {Peterson}\ \emph {et~al.}(2003)\citenamefont
  {Peterson}, \citenamefont {Figgen}, \citenamefont {Goll}, \citenamefont
  {Stoll},\ and\ \citenamefont
  {Dolg}}]{Peterson.Figgen.ea:Systematically.2003}%
  \BibitemOpen
  \bibfield  {author} {\bibinfo {author} {\bibfnamefont {K.~A.}\ \bibnamefont
  {Peterson}}, \bibinfo {author} {\bibfnamefont {D.}~\bibnamefont {Figgen}},
  \bibinfo {author} {\bibfnamefont {E.}~\bibnamefont {Goll}}, \bibinfo {author}
  {\bibfnamefont {H.}~\bibnamefont {Stoll}}, \ and\ \bibinfo {author}
  {\bibfnamefont {M.}~\bibnamefont {Dolg}},\ }\href {\doibase
  10.1063/1.1622924} {\bibfield  {journal} {\bibinfo  {journal} {J. Chem.
  Phys.}\ }\textbf {\bibinfo {volume} {119}},\ \bibinfo {pages} {11113}
  (\bibinfo {year} {2003})}\BibitemShut {NoStop}%
\bibitem [{\citenamefont {Peterson}\ \emph {et~al.}(2007)\citenamefont
  {Peterson}, \citenamefont {Figgen}, \citenamefont {Dolg},\ and\ \citenamefont
  {Stoll}}]{Peterson.Figgen.ea:Energy-consistent.2007}%
  \BibitemOpen
  \bibfield  {author} {\bibinfo {author} {\bibfnamefont {K.~A.}\ \bibnamefont
  {Peterson}}, \bibinfo {author} {\bibfnamefont {D.}~\bibnamefont {Figgen}},
  \bibinfo {author} {\bibfnamefont {M.}~\bibnamefont {Dolg}}, \ and\ \bibinfo
  {author} {\bibfnamefont {H.}~\bibnamefont {Stoll}},\ }\href {\doibase
  10.1063/1.2647019} {\bibfield  {journal} {\bibinfo  {journal} {J. Chem.
  Phys.}\ }\textbf {\bibinfo {volume} {126}},\ \bibinfo {pages} {124101}
  (\bibinfo {year} {2007})}\BibitemShut {NoStop}%
\bibitem [{\citenamefont {Mir\'o}, \citenamefont {Audiffred},\ and\
  \citenamefont {Heine}(2014)}]{Miro.Audiffred.ea:atlas.2014}%
  \BibitemOpen
  \bibfield  {author} {\bibinfo {author} {\bibfnamefont {P.}~\bibnamefont
  {Mir\'o}}, \bibinfo {author} {\bibfnamefont {M.}~\bibnamefont {Audiffred}}, \
  and\ \bibinfo {author} {\bibfnamefont {T.}~\bibnamefont {Heine}},\ }\href
  {\doibase 10.1039/C4CS00102H} {\bibfield  {journal} {\bibinfo  {journal}
  {Chem. Soc. Rev.}\ }\textbf {\bibinfo {volume} {43}},\ \bibinfo {pages}
  {6537} (\bibinfo {year} {2014})}\BibitemShut {NoStop}%
\bibitem [{\citenamefont {Perdew}\ \emph {et~al.}(2009)\citenamefont {Perdew},
  \citenamefont {Ruzsinszky}, \citenamefont {Csonka}, \citenamefont
  {Constantin},\ and\ \citenamefont
  {Sun}}]{Perdew.Ruzsinszky.ea:Workhorse.2009}%
  \BibitemOpen
  \bibfield  {author} {\bibinfo {author} {\bibfnamefont {J.~P.}\ \bibnamefont
  {Perdew}}, \bibinfo {author} {\bibfnamefont {A.}~\bibnamefont {Ruzsinszky}},
  \bibinfo {author} {\bibfnamefont {G.~I.}\ \bibnamefont {Csonka}}, \bibinfo
  {author} {\bibfnamefont {L.~A.}\ \bibnamefont {Constantin}}, \ and\ \bibinfo
  {author} {\bibfnamefont {J.}~\bibnamefont {Sun}},\ }\href {\doibase
  10.1103/PhysRevLett.103.026403} {\bibfield  {journal} {\bibinfo  {journal}
  {Phys. Rev. Lett.}\ }\textbf {\bibinfo {volume} {103}},\ \bibinfo {pages}
  {026403} (\bibinfo {year} {2009})}\BibitemShut {NoStop}%
\bibitem [{\citenamefont {Perdew}\ \emph {et~al.}(2011)\citenamefont {Perdew},
  \citenamefont {Ruzsinszky}, \citenamefont {Csonka}, \citenamefont
  {Constantin},\ and\ \citenamefont {Sun}}]{Perdew.Ruzsinszky.ea:Erratum.2011}%
  \BibitemOpen
  \bibfield  {author} {\bibinfo {author} {\bibfnamefont {J.~P.}\ \bibnamefont
  {Perdew}}, \bibinfo {author} {\bibfnamefont {A.}~\bibnamefont {Ruzsinszky}},
  \bibinfo {author} {\bibfnamefont {G.~I.}\ \bibnamefont {Csonka}}, \bibinfo
  {author} {\bibfnamefont {L.~A.}\ \bibnamefont {Constantin}}, \ and\ \bibinfo
  {author} {\bibfnamefont {J.}~\bibnamefont {Sun}},\ }\href {\doibase
  10.1103/PhysRevLett.106.179902} {\bibfield  {journal} {\bibinfo  {journal}
  {Phys. Rev. Lett.}\ }\textbf {\bibinfo {volume} {106}},\ \bibinfo {pages}
  {179902} (\bibinfo {year} {2011})}\BibitemShut {NoStop}%
\bibitem [{\citenamefont {Figgen}\ \emph {et~al.}(2005)\citenamefont {Figgen},
  \citenamefont {Rauhut}, \citenamefont {Dolg},\ and\ \citenamefont
  {Stoll}}]{Figgen.Rauhut.ea:Energy-consistent.2005}%
  \BibitemOpen
  \bibfield  {author} {\bibinfo {author} {\bibfnamefont {D.}~\bibnamefont
  {Figgen}}, \bibinfo {author} {\bibfnamefont {G.}~\bibnamefont {Rauhut}},
  \bibinfo {author} {\bibfnamefont {M.}~\bibnamefont {Dolg}}, \ and\ \bibinfo
  {author} {\bibfnamefont {H.}~\bibnamefont {Stoll}},\ }\href {\doibase
  https://doi.org/10.1016/j.chemphys.2004.10.005} {\bibfield  {journal}
  {\bibinfo  {journal} {Chem. Phys.}\ }\textbf {\bibinfo {volume} {311}},\
  \bibinfo {pages} {227} (\bibinfo {year} {2005})}\BibitemShut {NoStop}%
\bibitem [{\citenamefont {Peterson}\ \emph {et~al.}(2006)\citenamefont
  {Peterson}, \citenamefont {Shepler}, \citenamefont {Figgen},\ and\
  \citenamefont {Stoll}}]{Peterson.Shepler.ea:On.2006}%
  \BibitemOpen
  \bibfield  {author} {\bibinfo {author} {\bibfnamefont {K.~A.}\ \bibnamefont
  {Peterson}}, \bibinfo {author} {\bibfnamefont {B.~C.}\ \bibnamefont
  {Shepler}}, \bibinfo {author} {\bibfnamefont {D.}~\bibnamefont {Figgen}}, \
  and\ \bibinfo {author} {\bibfnamefont {H.}~\bibnamefont {Stoll}},\ }\href
  {\doibase 10.1021/jp065887l} {\bibfield  {journal} {\bibinfo  {journal} {J.
  Phys. Chem. A}\ }\textbf {\bibinfo {volume} {110}},\ \bibinfo {pages} {13877}
  (\bibinfo {year} {2006})}\BibitemShut {NoStop}%
\bibitem [{\citenamefont {Liu}, \citenamefont {Feng},\ and\ \citenamefont
  {Yao}(2011)}]{Liu.Feng.ea:Quantum.2011}%
  \BibitemOpen
  \bibfield  {author} {\bibinfo {author} {\bibfnamefont {C.-C.}\ \bibnamefont
  {Liu}}, \bibinfo {author} {\bibfnamefont {W.}~\bibnamefont {Feng}}, \ and\
  \bibinfo {author} {\bibfnamefont {Y.}~\bibnamefont {Yao}},\ }\href {\doibase
  10.1103/PhysRevLett.107.076802} {\bibfield  {journal} {\bibinfo  {journal}
  {Phys. Rev. Lett.}\ }\textbf {\bibinfo {volume} {107}},\ \bibinfo {pages}
  {076802} (\bibinfo {year} {2011})}\BibitemShut {NoStop}%
\bibitem [{\citenamefont {Kang}\ \emph {et~al.}(2019)\citenamefont {Kang},
  \citenamefont {Li}, \citenamefont {Sohn}, \citenamefont {Shan},\ and\
  \citenamefont {Mak}}]{Kang.Li.ea:Nonlinear.2019}%
  \BibitemOpen
  \bibfield  {author} {\bibinfo {author} {\bibfnamefont {K.}~\bibnamefont
  {Kang}}, \bibinfo {author} {\bibfnamefont {T.}~\bibnamefont {Li}}, \bibinfo
  {author} {\bibfnamefont {E.}~\bibnamefont {Sohn}}, \bibinfo {author}
  {\bibfnamefont {J.}~\bibnamefont {Shan}}, \ and\ \bibinfo {author}
  {\bibfnamefont {K.~F.}\ \bibnamefont {Mak}},\ }\href {\doibase
  10.1038/s41563-019-0294-7} {\bibfield  {journal} {\bibinfo  {journal} {Nature
  Materials}\ }\textbf {\bibinfo {volume} {18}},\ \bibinfo {pages} {324}
  (\bibinfo {year} {2019})}\BibitemShut {NoStop}%
\bibitem [{\citenamefont {Zhu}, \citenamefont {Cheng},\ and\ \citenamefont
  {Schwingenschl\"ogl}(2011)}]{Zhu.Cheng.ea:Giant.2011}%
  \BibitemOpen
  \bibfield  {author} {\bibinfo {author} {\bibfnamefont {Z.~Y.}\ \bibnamefont
  {Zhu}}, \bibinfo {author} {\bibfnamefont {Y.~C.}\ \bibnamefont {Cheng}}, \
  and\ \bibinfo {author} {\bibfnamefont {U.}~\bibnamefont
  {Schwingenschl\"ogl}},\ }\href {\doibase 10.1103/PhysRevB.84.153402}
  {\bibfield  {journal} {\bibinfo  {journal} {Phys. Rev. B}\ }\textbf {\bibinfo
  {volume} {84}},\ \bibinfo {pages} {153402} (\bibinfo {year}
  {2011})}\BibitemShut {NoStop}%
\bibitem [{\citenamefont {Kadek}\ \emph {et~al.}(2023)\citenamefont {Kadek},
  \citenamefont {Wang}, \citenamefont {Joosten}, \citenamefont {Chiu},
  \citenamefont {Mairesse}, \citenamefont {Repisky}, \citenamefont {Ruud},\
  and\ \citenamefont {Bansil}}]{Kadek.Wang.ea:Band.2023}%
  \BibitemOpen
  \bibfield  {author} {\bibinfo {author} {\bibfnamefont {M.}~\bibnamefont
  {Kadek}}, \bibinfo {author} {\bibfnamefont {B.}~\bibnamefont {Wang}},
  \bibinfo {author} {\bibfnamefont {M.}~\bibnamefont {Joosten}}, \bibinfo
  {author} {\bibfnamefont {W.-C.}\ \bibnamefont {Chiu}}, \bibinfo {author}
  {\bibfnamefont {F.}~\bibnamefont {Mairesse}}, \bibinfo {author}
  {\bibfnamefont {M.}~\bibnamefont {Repisky}}, \bibinfo {author} {\bibfnamefont
  {K.}~\bibnamefont {Ruud}}, \ and\ \bibinfo {author} {\bibfnamefont
  {A.}~\bibnamefont {Bansil}},\ }\href {\doibase
  10.1103/PhysRevMaterials.7.064001} {\bibfield  {journal} {\bibinfo  {journal}
  {Phys. Rev. Mater.}\ }\textbf {\bibinfo {volume} {7}},\ \bibinfo {pages}
  {064001} (\bibinfo {year} {2023})}\BibitemShut {NoStop}%
\bibitem [{\citenamefont {Franzke}\ and\ \citenamefont
  {Holzer}(2022)}]{Franzke.Holzer:Impact.2022}%
  \BibitemOpen
  \bibfield  {author} {\bibinfo {author} {\bibfnamefont {Y.~J.}\ \bibnamefont
  {Franzke}}\ and\ \bibinfo {author} {\bibfnamefont {C.}~\bibnamefont
  {Holzer}},\ }\href {\doibase 10.1063/5.0103898} {\bibfield  {journal}
  {\bibinfo  {journal} {J. Chem. Phys.}\ }\textbf {\bibinfo {volume} {157}},\
  \bibinfo {pages} {031102} (\bibinfo {year} {2022})}\BibitemShut {NoStop}%
\bibitem [{\citenamefont {Grotjahn}, \citenamefont {Furche},\ and\
  \citenamefont {Kaupp}(2022)}]{Grotjahn.Furche.ea:Importance.2022}%
  \BibitemOpen
  \bibfield  {author} {\bibinfo {author} {\bibfnamefont {R.}~\bibnamefont
  {Grotjahn}}, \bibinfo {author} {\bibfnamefont {F.}~\bibnamefont {Furche}}, \
  and\ \bibinfo {author} {\bibfnamefont {M.}~\bibnamefont {Kaupp}},\ }\href
  {\doibase 10.1063/5.0113083} {\bibfield  {journal} {\bibinfo  {journal} {J.
  Chem. Phys.}\ }\textbf {\bibinfo {volume} {157}},\ \bibinfo {pages} {111102}
  (\bibinfo {year} {2022})}\BibitemShut {NoStop}%
\bibitem [{\citenamefont {Grimme}\ \emph {et~al.}(2010)\citenamefont {Grimme},
  \citenamefont {Antony}, \citenamefont {Ehrlich},\ and\ \citenamefont
  {Krieg}}]{Grimme.Antony.ea:consistent.2010}%
  \BibitemOpen
  \bibfield  {author} {\bibinfo {author} {\bibfnamefont {S.}~\bibnamefont
  {Grimme}}, \bibinfo {author} {\bibfnamefont {J.}~\bibnamefont {Antony}},
  \bibinfo {author} {\bibfnamefont {S.}~\bibnamefont {Ehrlich}}, \ and\
  \bibinfo {author} {\bibfnamefont {H.}~\bibnamefont {Krieg}},\ }\href
  {\doibase http://dx.doi.org/10.1063/1.3382344} {\bibfield  {journal}
  {\bibinfo  {journal} {J. Chem. Phys.}\ }\textbf {\bibinfo {volume} {132}},\
  \bibinfo {pages} {154104} (\bibinfo {year} {2010})}\BibitemShut {NoStop}%
\bibitem [{\citenamefont {Grimme}, \citenamefont {Ehrlich},\ and\ \citenamefont
  {Goerigk}(2011)}]{Grimme.Ehrlich.ea:Effect.2011}%
  \BibitemOpen
  \bibfield  {author} {\bibinfo {author} {\bibfnamefont {S.}~\bibnamefont
  {Grimme}}, \bibinfo {author} {\bibfnamefont {S.}~\bibnamefont {Ehrlich}}, \
  and\ \bibinfo {author} {\bibfnamefont {L.}~\bibnamefont {Goerigk}},\ }\href
  {\doibase 10.1002/jcc.21759} {\bibfield  {journal} {\bibinfo  {journal} {J.
  Comput. Chem.}\ }\textbf {\bibinfo {volume} {32}},\ \bibinfo {pages} {1456}
  (\bibinfo {year} {2011})}\BibitemShut {NoStop}%
\bibitem [{\citenamefont {Patra}\ \emph {et~al.}(2020)\citenamefont {Patra},
  \citenamefont {Jana}, \citenamefont {Constantin},\ and\ \citenamefont
  {Samal}}]{Patra.Jana.ea:Efficient.2020}%
  \BibitemOpen
  \bibfield  {author} {\bibinfo {author} {\bibfnamefont {A.}~\bibnamefont
  {Patra}}, \bibinfo {author} {\bibfnamefont {S.}~\bibnamefont {Jana}},
  \bibinfo {author} {\bibfnamefont {L.~A.}\ \bibnamefont {Constantin}}, \ and\
  \bibinfo {author} {\bibfnamefont {P.}~\bibnamefont {Samal}},\ }\href
  {\doibase 10.1063/5.0011849} {\bibfield  {journal} {\bibinfo  {journal} {J.
  Chem. Phys.}\ }\textbf {\bibinfo {volume} {153}},\ \bibinfo {pages} {084117}
  (\bibinfo {year} {2020})}\BibitemShut {NoStop}%
\bibitem [{\citenamefont {Ehlert}\ \emph {et~al.}(2021)\citenamefont {Ehlert},
  \citenamefont {Huniar}, \citenamefont {Ning}, \citenamefont {Furness},
  \citenamefont {Sun}, \citenamefont {Kaplan}, \citenamefont {Perdew},\ and\
  \citenamefont {Brandenburg}}]{Ehlert.Huniar.ea:r2SCAN-D4.2021}%
  \BibitemOpen
  \bibfield  {author} {\bibinfo {author} {\bibfnamefont {S.}~\bibnamefont
  {Ehlert}}, \bibinfo {author} {\bibfnamefont {U.}~\bibnamefont {Huniar}},
  \bibinfo {author} {\bibfnamefont {J.}~\bibnamefont {Ning}}, \bibinfo {author}
  {\bibfnamefont {J.~W.}\ \bibnamefont {Furness}}, \bibinfo {author}
  {\bibfnamefont {J.}~\bibnamefont {Sun}}, \bibinfo {author} {\bibfnamefont
  {A.~D.}\ \bibnamefont {Kaplan}}, \bibinfo {author} {\bibfnamefont {J.~P.}\
  \bibnamefont {Perdew}}, \ and\ \bibinfo {author} {\bibfnamefont {J.~G.}\
  \bibnamefont {Brandenburg}},\ }\href {\doibase 10.1063/5.0041008} {\bibfield
  {journal} {\bibinfo  {journal} {J. Chem. Phys.}\ }\textbf {\bibinfo {volume}
  {154}},\ \bibinfo {pages} {061101} (\bibinfo {year} {2021})}\BibitemShut
  {NoStop}%
\end{thebibliography}%

\ifarXiv
\foreach \x in {1,...,\numbersupplementpages}
{
	\clearpage
	\includepdf[pages={\x,{}}]{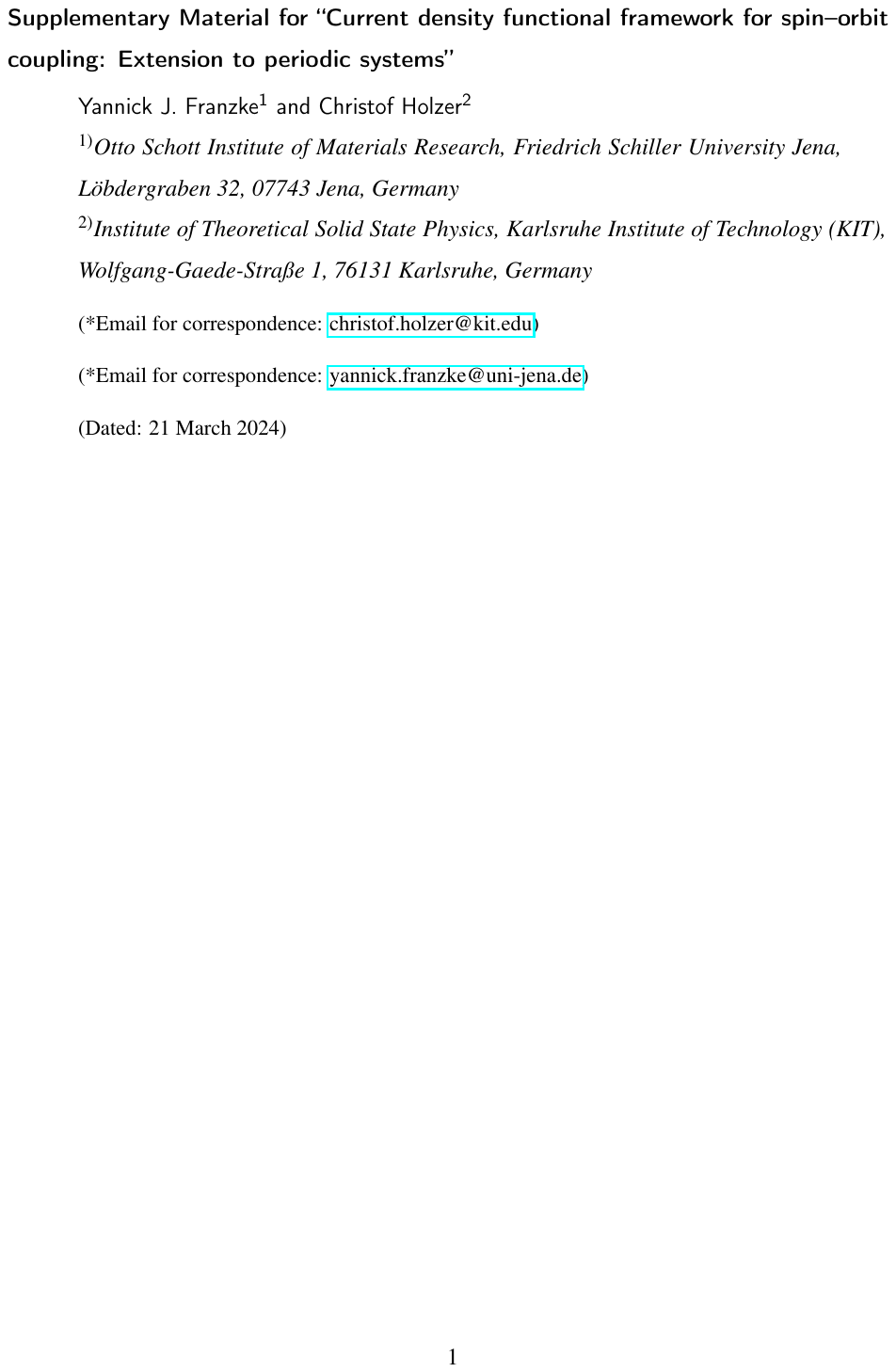}
}
\fi

\end{document}